\documentclass[twocolumn,showpacs,preprintnumbers,amsmath,amssymb,prl,longbibliography,superscriptaddress]{revtex4-1}
\usepackage{bbm}
\usepackage{mathrsfs}
\usepackage{graphicx}
\usepackage{dcolumn}
\usepackage{bm}
\usepackage{amsmath}
\usepackage{amsfonts}
\usepackage{color}
\usepackage[colorlinks=true,citecolor=blue,anchorcolor=blue]{hyperref}
\usepackage{floatrow}

\begin{document}

\title{Intrinsic topological phases in Mn$_2$Bi$_2$Te$_5$ tuned by the layer magnetization}
\author{Yue Li}
\affiliation{State Key Laboratory of Surface Physics and Department of Physics, Fudan University, Shanghai 200433, China}
\author{Yadong Jiang}
\affiliation{State Key Laboratory of Surface Physics and Department of Physics, Fudan University, Shanghai 200433, China}
\author{Jinlong Zhang}
\affiliation{State Key Laboratory of Surface Physics and Department of Physics, Fudan University, Shanghai 200433, China}
\author{Zhaochen Liu}
\affiliation{State Key Laboratory of Surface Physics and Department of Physics, Fudan University, Shanghai 200433, China}
\author{Zhongqin Yang}
\affiliation{State Key Laboratory of Surface Physics and Department of Physics, Fudan University, Shanghai 200433, China}
\affiliation{Key Laboratory of Computational Physical Sciences (Ministry of Education), Fudan University, Shanghai 200433, China}
\author{Jing Wang}
\thanks{To whom correspondence should be addressed.\\ wjingphys@fudan.edu.cn}
\affiliation{State Key Laboratory of Surface Physics and Department of Physics, Fudan University, Shanghai 200433, China}
\affiliation{Institute for Nanoelectronic Devices and Quantum Computing, Fudan University, Shanghai 200433, China}

\begin{abstract}
The interplay between band topology and magnetic order could generate a variety of time-reversal-breaking gapped topological phases with exotic topological quantization phenomena, such as quantum anomalous Hall (QAH) insulators and axion insulators (AxI). Here by combining analytic models and first-principles calculations, we find extremely rich magnetic topological quantum states in a van der Waals layered material Mn$_2$Bi$_2$Te$_5$, including a dynamic axion field in the antiferromagnetic bulk, an ideal magnetic Weyl semimetal with a single pair of Weyl points, as well as QAH and AxI phases in thin films. The phase transition between QAH and AxI is tuned by the layer magnetization, which would provide a promising platform for chiral superconducting phases and Majorana fermion. We further present a simple and unified continuum model that captures the magnetic topological features, and is generic for Mn$_2$Bi$_2$Te$_5$ and MnBi$_2$Te$_4$ family materials.
\end{abstract}

\date{\today}


\maketitle

The discovery of time-reversal-invariant topological insulator brings the opportunity to realize a large family of exotic topological quantization phenomena~\cite{kane2005,kane2005b,bernevig2006c,koenig2007,fu2007,hasan2010,qi2011,tokura2019,wang2017c}. The interplay between band topology and magnetism could give rise to a variety of exotic time-reversal-breaking gapped topological states, including the quantum anomalous Hall (QAH) effect with dissipationless chiral edge states~\cite{haldane1988,qi2006,liu2008,yu2010,wang2013a,wang2015d,liu2016,qi2008}, axion insulator (AxI) displaying topological magnetoelectric effects~\cite{qi2008,essin2009,li2010,wan2012,wang2015b,morimoto2015,wang2016a,mogi2017,xu2019,chowdhury2019}, and chiral superconducting state with Majorana fermions (if in proximity to superconductors)~\cite{read2000,alicea2012,qi2010b,wang2015c}. Interestingly, the QAH effect may find applications in low-power-consumption electronics and non-Abelian braiding of Majorana fermions is useful in topological computation~\cite{ivanov2001,kitaev2003,nayak2008,alicea2011,lian2018b}. Despite of these predicted important physical effects, until now only a few of them have been experimentally proved, due to a limited number of magnetic topological insulator (TI) materials. A prime example is the QAH effect experimentally observed in magnetically doped (Bi,Sb)$_2$Te$_3$ film~\cite{chang2013b,checkelsky2014,bestwick2015,chang2015}. However, the random magnetic dopants limit the quality and exchange gap~\cite{lee2015} of the material, which further constrain the quantization of AHE appearing only at very low temperatures. In proximity with an $s$-wave superconductor, such a strongly disordered QAH system at coercivity by the random magnetic domains complicates the transport experiments in millimeter-size sample~\cite{he2017,kayyalha2020,ji2018,huang2018,lian2018}. Therefore, finding stoichiometric TI with an innate magnetic order are highly desired, which would provide a homogenous platform for high temperature QAH effect and coherent chiral Majorana fermions.

The first intrinsic magnetic TI MnBi$_2$Te$_4$ (MBT) discovered recently is an interesting candidate for observing these topological phenomena~\cite{zhang2019,li2019,gong2019,otrokov2019,deng2020,liu2020,ge2020,lee2019,yan2019,hao2019,lih2019,chen2019}. For instance, the zero-field QAH effect has been observed at an elevated temperature~\cite{deng2020}. Given the importance of magnetic TIs as a platform for new states of quantum matter, it is important to search for \emph{new} material systems that are stoichiometric crystals with well-defined electronic structures, preferably with simple surface states, and describable by simple theoretical models. In this work, by combining analytic models and first-principles calculations, we predict rich topological quantum states in new magnetic TI family Mn$_2$Bi$_2$Te$_5$ (M$_2$BT). The antiferromagnetic (AFM) bulk hosts a dynamical axion field, which was first proposed in Ref.~\cite{zhang2020} focusing on the interplay of bulk AFM fluctuations and axion electrodynamics. Here we will focus on the various topological states in its bulk and thin film forms with different \emph{static} magnetic ordering.

\begin{figure}[t]
\begin{center}
\includegraphics[width=3.4in,clip=true]{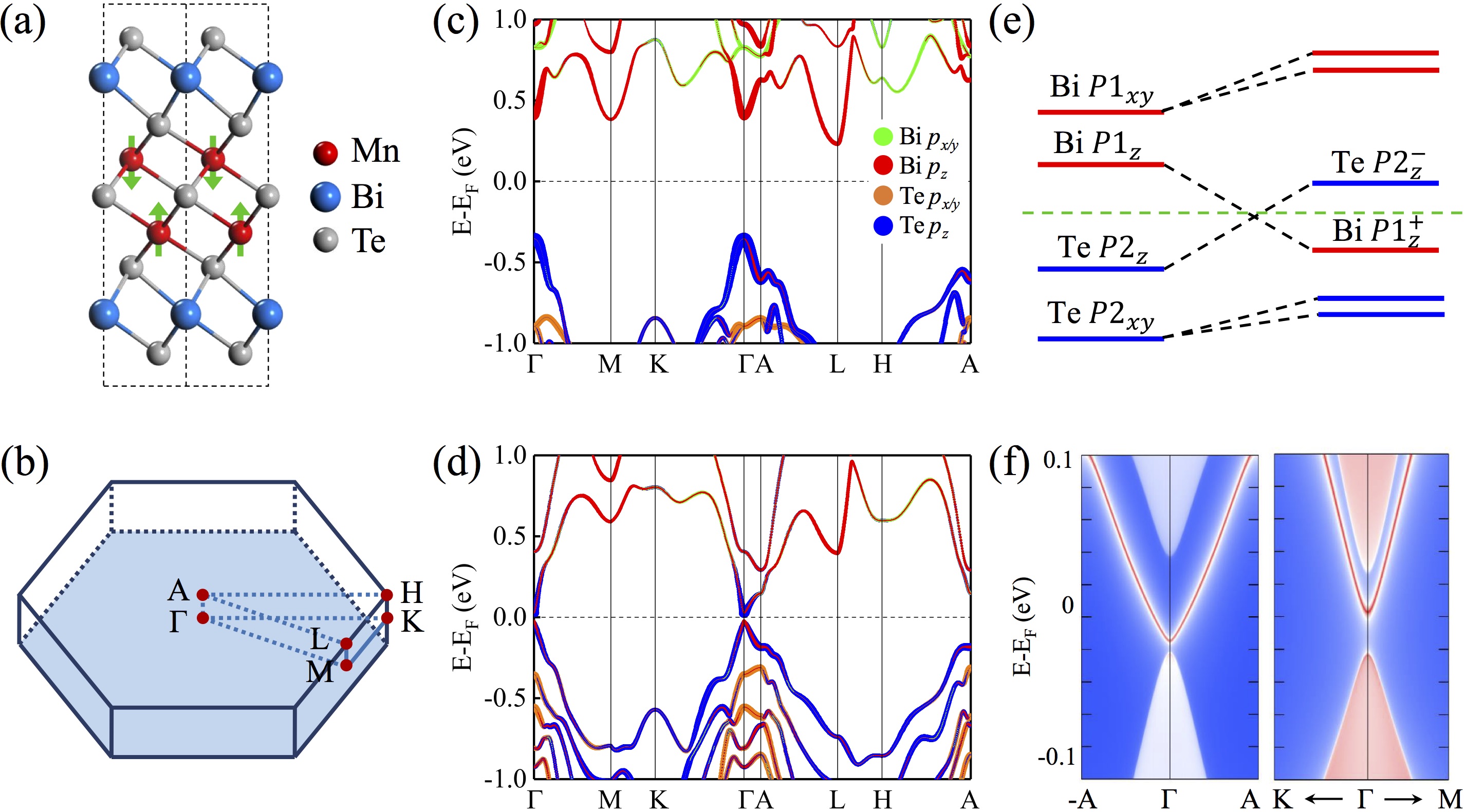}
\end{center}
\caption{AFM bulk M$_2$BT. (a) The lattice structure with an $A$-type AFM ordering. The black dotted lines indicate the unit cell as 1 NL. The green arrows represent the spin moments of Mn atoms. (b) Bulk Brillouin zone. (c) and (d) give the orbital-projected band structures without and with SOC, respectively. (e) Schematic diagram of the band inversion induced by SOC at $\Gamma$. The green dashed line represents the Fermi level. (f) The energy and momentum dependence of the local density of states (LDOS) on the ($1\bar{1}0$) and ($001$) surface, respectively.}
\label{fig1}
\end{figure}

M$_2$BT is a layered ternary tetradymite compound that consists of ABC stacking Te$1$-Bi$1$-Te$2$-Mn$1$-Te$3$-Mn$1'$-Te$2'$-Bi$1'$-Te$1'$ nonuple layers (NL), which has been successfully synthesized in experiments recently~\cite{lv2020}. It has a hexagonal crystal structure shown in Fig.~\ref{fig1}(a) with space group $P\bar{3}m1$ (No.~164), which can be viewed as layered TI Bi$_2$Te$_3$ with each of its Te-Bi-Te-Bi-Te quintuple layer intercalated by two additional Mn-Te bilayers. The trigonal axis (threefold rotation symmetry $C_{3z}$) is defined as the $z$ axis, a binary axis (twofold rotation symmetry $C_{2x}$) is defined as the $x$ axis and a bisectrix axis (in the reflection plane) is defined as the $y$ axis for the coordinate system. 
The system has inversion symmetry $\mathcal{P}$ with Te$3$ site as an inversion center if the spin moments of Mn are ignored.

As far as the magnetic order is concerned, it appears that the Mn spins couple ferromagnetically in each layer, but the adjacent Mn layers couple anti-parallel to each other. The ferromagnetic (FM) order in each Mn layer can be understood from the Goodenough-Kanamori $90^\circ$ rule, while the AFM coupling between adjacent Mn layers is from the interlayer superexchange similar to MBT. The local magnetic moments are roughly $4.59\mu_B$ independent of the film thickness. Table~\ref{table} lists the thickness dependence of magnetism, and the magnetic anisotropic energy (MAE) for 2 to 7 NL as well as bulk are about 0.1~meV/Mn and insensitive to layer thickness, indicating the N\'eel-type AFM order along $z$ axis is the ground state. The non-collinear and other possible collinear magnetic configurations are found to have higher energies~\cite{hou2019}. Therefore for the AFM state in both bulk and film, the time reversal ($\mathcal{T}$) and $\mathcal{P}$ are broken, but $\mathcal{PT}$ is conserved. This is in sharp contrast to AFM MBT, where its even layer film breaks $\mathcal{T}$ and $\mathcal{P}$; while its bulk conserves $\mathcal{P}$ and $\mathcal{T}\tau_{1/2}$, in which $\tau_{1/2}$ is the half-translation operator along $z$ axis. Thus, a $Z_2$ invariant is well defined for bulk MBT as a AFM TI with quantized axion response ($\theta=\pi$ defined module $2\pi$ as in Lagrangian $\mathcal{L}_{\theta}=(\theta/2\pi)(e^2/h)\mathbf{E}\cdot\mathbf{B}$~\cite{qi2008}), while bulk M$_2$BT is a magnetic insulator but with a nonquantized $\theta$ response. 

\begin{table}[b]
\caption{Thickness dependence of M$_2$BT films magnetism and the energy gap in AFM and FM states. $\Delta E_{\text{A/F}}=E_{\text{AFM}}-E_{\text{FM}}$ is the total energy difference of the AFM and FM states along $z$ direction. The N\'{e}el type AFM is the ground state. The AFM thin films are AxI with $\mathcal{C}=0$; while for the FM films, $\mathcal{C}=0$ for 1 NL and $\mathcal{C}=1$ for 2-7 NL from first-principles calculations.}
\begin{center}\label{table}
\renewcommand{\arraystretch}{1.6}
\begin{tabular*}{3.4in}
{@{\extracolsep{\fill}}ccccc}
\hline
\hline
Thickness & $\Delta E_{\text{A/F}}$ & MAE & $E_g$ (AFM) & $E_g$ (FM)
\\
(NL) & [meV/Mn] & [meV/Mn] & [meV] & [meV] 
\\
\hline
1 & $-3.726$ & $0.052$ & $407.9$ & $83.6$
\\
2 &  $-3.877$ & $0.094$ & $67.1$ & $43.1$
\\
3 &  $-3.875$ & $0.108$ & $29.5$ & $60.7$
\\
4 &  $-3.643$ & $0.104$ & $31.0$ & $52.8$
\\
5 &  $-3.614$ & $0.109$ & $20.0$ & $41.2$
\\
6 &  $-3.789$ & $0.110$ & $24.5$ & $31.1$
\\
7 &  $-3.740$ & $0.111$ & $19.4$ & $19.3$
\\
$\infty$ (bulk) &  $-4.344$ & $0.117$ & $50.6$ & $0$
\\
\hline
\hline
\end{tabular*}
\end{center}
\end{table}

Then we turn to the electronic and topological properties of the material. To have an intuitive understanding of the underline physics, we start with the bulk electronic structure. The detailed methods can be found in the Supplemental Material~\cite{supplementary}. For the AFM ground state, the band structures without and with spin-orbit coupling (SOC) are shown in Figs.~\ref{fig1}(c) and~\ref{fig1}(d), respectively. Mn $d$-bands are far away from the band gap due to a large spin splitting ($>5$~eV), and only Bi/Te $p_z$-bands are close to the Fermi level with an anticrossing feature around the $\Gamma$ point from the band inversion, suggesting the nontrivial topology in bulk M$_2$BT. To characterize the low-energy physics, an effective model is constructed~\cite{zhang2020}. As shown in Fig.~\ref{fig1}(e), the low-lying states at $\Gamma$ are the $|P1_z^+\rangle$ of two Bi layers and $|P2_z^-\rangle$ of two Te layers (Te$1$ and Te$1'$), the superscripts ``$+$'', ``$-$'' stand for parity. The SOC further leads to band inversion. The symmetries of AFM system are the three-fold rotation symmetry $C_{3z}$ and $\mathcal{PT}$. In the basis of $(|P1^+_z,\uparrow\rangle,|P1^+_z,\downarrow\rangle,|P2^-_z,\uparrow\rangle,|P2^-_z,\downarrow\rangle)$, the representation of the symmetry operations is given by $C_{3z}=\exp[i(\pi/3)\sigma^z\otimes1]$ and $\mathcal{PT}=i\sigma^y\mathcal{K}\otimes\tau^z$ ($\mathcal{T}=i\sigma^y\mathcal{K}\otimes1$, $\mathcal{P}=1\otimes\tau^z$), where $\mathcal{K}$ is complex conjugation operator, $\sigma^{x,y,z}$ and $\tau^{x,y,z}$ denote the Pauli matrices in the spin and orbital space, respectively. The generic form of the AFM Hamiltonian obeying these symmetries is
\begin{eqnarray}\label{AFM}
\mathcal{H}_{\text{AFM}}(\mathbf{k}) &=& A_1k_z\sigma^z\otimes\tau^x+\mathcal{A}_2(k_z)(k_y\sigma^x-k_x\sigma^y)\otimes\tau^x
\nonumber
\\
&+&M_4(\mathbf{k})1\otimes\tau^z+M_5(\mathbf{k})1\otimes\tau^y+\epsilon_0(\mathbf{k}),
\end{eqnarray}
where $\epsilon_0(\mathbf k)=C_0+C_1k_z^2+C_2(k_x^2+k_y^2)$, $M_4(\mathbf k)=M_0+M_1k_z^2+M_2(k_x^2+k_y^2)$, $M_5(\mathbf{k})=B_0+B_1k_z^2+B_2(k_x^2+k_y^2)$, and $\mathcal{A}_2(k_z)=A_2+A_3k_z$. Here $M_0<0$ and $M_{1,2}>0$ correctly characterizes the band inversion at $\Gamma$~\cite{supplementary}. Without $M_5(\mathbf{k})$ and $A_3$ terms, Eq.~(\ref{AFM}) is nothing but the textbook TI model in Bi$_2$Te$_3$ family with a single surface Dirac cone~\cite{zhang2009}. $M_5(\mathbf{k})$ and $A_3$ are $\mathcal{T}, \mathcal{P}$-breaking perturbations induced by the $z$-direction N\'eel order on Mn. The direct consequence of $M_5(\mathbf{k})$ term is to open a gap in the surface-state spectrum with the sign independent of the surface orientation, which is equivalent to induce a hedgehodge magnetization on the TI surface. This is confirmed by the ($1\bar{1}0$) and ($001$) surface spectra by first-principles calculations in Fig.~\ref{fig1}(f), which are different from the gapless Dirac state on $\mathcal{T}\tau_{1/2}$-preserving surface in MBT. Also, they are different from the $x$ axis AFM state in M$_2$BT, which has gapless surface state on the surfaces parallel to N\'eel order with surface Dirac cone shifted away from $\Gamma$~\cite{supplementary}. The hedgehodge-like surface gap of AFM-$z$ M$_2$BT make it an ideal platform for topological magnetoelectric effect. The calculated static $\theta\approx0.83\pi$. Interestingly, $\theta$ becomes a dynamic axion field when the magnetic fluctuations are considered~\cite{li2010}.

\begin{figure}[t]
\begin{center}
\includegraphics[width=3.4in,clip=true]{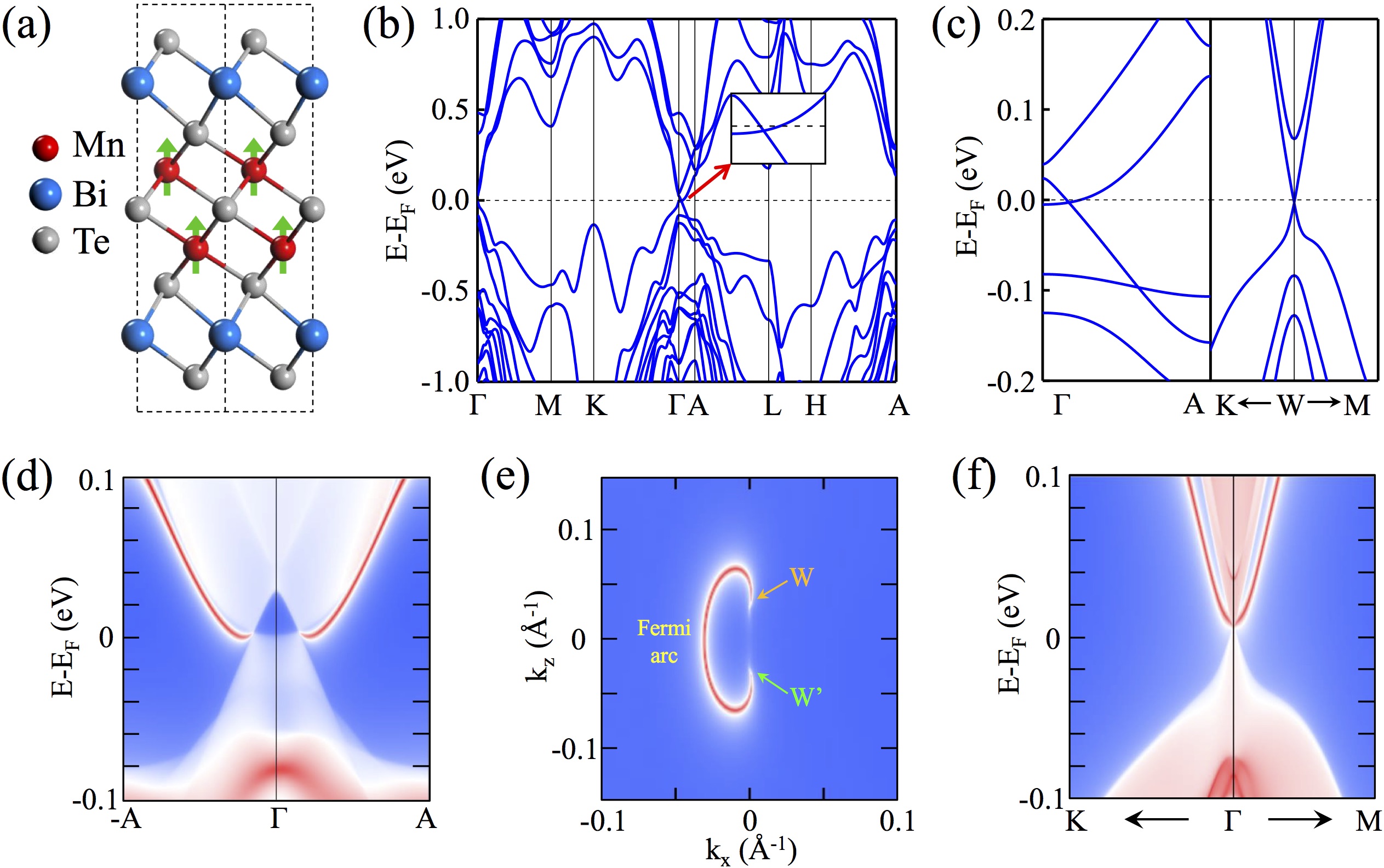}
\end{center}
\caption{FM bulk M$_2$BT. (a) Lattice structure. (b) The band structure with SOC. (c) Zoom-in band structures along the $\Gamma$-A and K-W-M directions. (d) \& (f) The energy and momentum dependence of the LDOS on the ($1\bar{1}0$) and (001) surfaces, respectively. The two Weyl points are seen along the A-$\Gamma$-A direction. (e) Surface states of the ($1\bar{1}0$) termination on the isoenergy plane of the Weyl points, demonstrating the existence of the Fermi arcs.}
\label{fig2}
\end{figure}

The AFM ground state of M$_2$BT could be tuned into the FM state by a magnetic field. From above we see low energy physics in AFM M$_2$BT is described by a TI model and $\mathcal{T,P}$-breaking perturbations. For a $z$-axis FM order, a $\mathcal{T}$-breaking but $\mathcal{P}$-conversing perturbation should be added, and the resulting possible phase is either Weyl semimetal, 3D QAH or trivial magnetic insulator~\cite{burkov2011,wang2016a}. The band structure of $z$-axis FM M$_2$BT bulk in Fig.~\ref{fig2} displays a pair of band crossings at Weyl points (W$'$ and W) along the $\bar{\text{A}}$-$\Gamma$-A line. The Wilson loop calculations suggest the Chern number $\mathcal{C}=1$ at $k_z=0$ plane and $\mathcal{C}=0$ at $k_z=\pi$ plane~\cite{supplementary}, which is consist with the minimal ideal Weyl semimetal in Fig.~\ref{fig2}(c). Figs.~\ref{fig2}(d)-(f) shows surface-state on different typical surfaces, where Fermi arcs on ($1\bar{1}0$) termination are clearly seen in Fig.~\ref{fig2}(e). Explicitly, the additional $\mathcal{T}$-breaking but $\mathcal{P}$-conversing terms describing the $z$-axis FM state is
\begin{eqnarray}\label{FM}
\delta\mathcal{H}_{\text{FM}}(\mathbf{k}) &=& A_3'k_z1\otimes\tau^x+A_2'(k_y\sigma^x-k_x\sigma^y)\otimes\tau^y
\nonumber
\\
&&+\ M_{1}^z(\mathbf{k})\sigma^z\otimes1+M_2^z(\mathbf{k})\sigma^z\otimes\tau^z,
\end{eqnarray}
where $M_{j}^z(\mathbf k)=D^j_0+D^j_1k_z^2+D^j_2(k_x^2+k_y^2)$ with $j=1,2$. This model is similar to FM MBT but with different parameters~\cite{supplementary}.

\begin{figure}[t]
\begin{center}
\includegraphics[width=3.4in,clip=true]{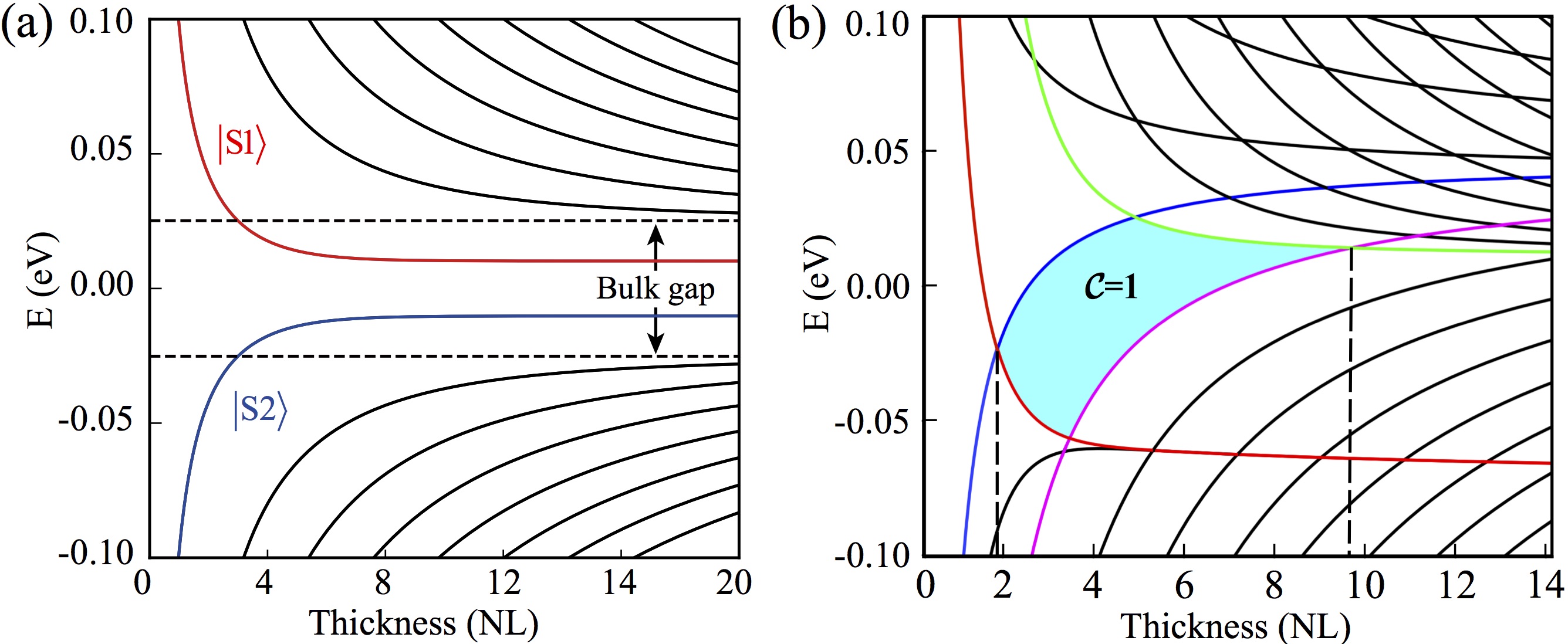}
\end{center}
\caption{The subbands energy level versus the thickness of the thin film for (a) AFM and (b) FM. In (a), the gap of the AFM film converges quickly as thickness exceeds $5$ NL. The density of $|S1\rangle$ and $|S2\rangle$ are localized on surfaces. In (b), the band inversion of first pair of polarized bands (red and blue lines which are localized on surfaces) leads to $\mathcal{C}=1$ in the shaded region. The second polarized band (green and purple lines) inversion suggests $\mathcal{C}=2$ when the film is $10$~NL or thicker. }
\label{fig3}
\end{figure}

Now we understand that the magnetic TI M$_2$BT is well described by a Bi$_2$Te$_3$-type TI model with corresponding $\mathcal{T}$-breaking perturbations introduced by Mn. The band inversion in 3D suggests the nontrivial topology may also exist in 2D, which we characterize in below. AFM M$_2$BT films have $\mathcal{PT}$ symmetry which leads to $\mathcal{C}=0$. They are magnetic insulator with nonquantized $\theta$ response from $\mathcal{P,T}$ breaking and finite-size effect. We calculate the energy level versus the film thickness of AFM Hamiltonian~(\ref{AFM}). Due to quantum confinement, the bulk bands become 2D subbands. As shown in Fig.~\ref{fig3}(a), the gap converges when the film exceeds $5$~NL, which is consistent with the first-principles calculations listed in Table~\ref{table}. The first pair of subbands $|S1\rangle$ and $|S2\rangle$ are localized on the two surfaces of the thin film~\cite{supplementary,liu2010a}, with a decay length of about $2$~NL. For FM films, $\mathcal{T}$ and $\mathcal{PT}$-breaking but $\mathcal{P}$-conversing leads to spin polarized bands, allowing $\mathcal{C}\neq0$. As calculated in Table~\ref{table}, $\mathcal{C}=0$ for $1$ NL and $\mathcal{C}=1$ for $2$-$7$ NL. The spin polarized energy level versus the film thickness in the FM state is calculated in Fig.~\ref{fig3}(b), and $\mathcal{C}$ is determined by the number of polarized band inversion~\cite{wang2013a}. Fig.~\ref{fig3}(b) suggests 3~NL has the maximum gap in $\mathcal{C}=1$ QAH and is consistent with first-principles calculations. 

\begin{figure}[t]
\begin{center}
\includegraphics[width=3.4in,clip=true]{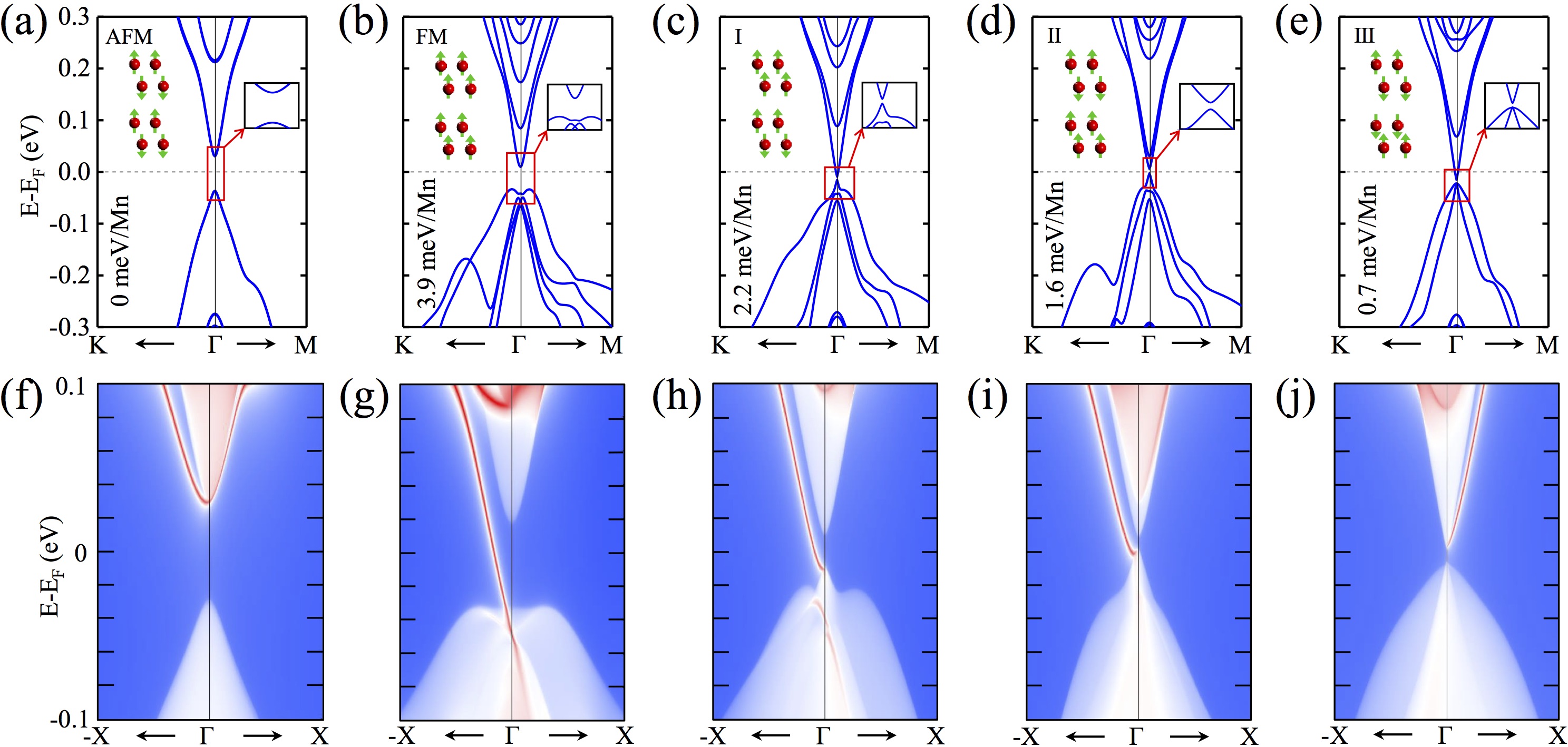}
\end{center}
\caption{2NL M$_2$BT film. (a-e) Band structures with AFM, FM, and interstate magnetic configurations (denoted as I, II, III). The relative total energies are shown, where the total energy of the reference AFM state is set to be zero. The band gap at $\Gamma$ point for (a-e) are $E_g=67.1, 51.5, 6.2, 8.7, 6.4$~meV. (f-j) The energy and momentum dependence of LDOS on the ($1\bar{1}$) edge with AFM, FM, I,II, and III orders. Interstate II can be obtained from the AFM state by applying a $z$-axis magnetic field.}
\label{fig4}
\end{figure}

Intriguingly, here as the interlayer exchange coupling is quite weak, the Mn layers may be driven into different magnetic configurations, which further modify the band topology. Take $2$ NL for example, we calculate the band structure, relative total energy and $\mathcal{C}$ for five different magnetic configurations named AFM, FM, interstate I, II, III shown in Fig.~\ref{fig4}. Clearly, FM, I and II are QAH with $\mathcal{C}=1$. AFM and III have $\mathcal{C}=0$, lead to zero Hall conductance. As we show below, the AFM state is a magnetic insulator with nonquantized $\theta$ but III is trivial insulator. 

To describe the layer magnetization tuned QAH state in M$_2$BT film, we start with the low energy physics which is well described by the massive Dirac surface states only, where the intrinsic magnetic ordering introduces different Zeeman terms on these two surfaces. The generic effective Hamiltonian for thin film is
\begin{eqnarray}\label{afm}
\mathcal{H}_{\text{film}}(k_x,k_y) &=& v_F(k_y\sigma_x-k_x\sigma_y)\otimes\tau_z+m(k)1\otimes\tau_x
\nonumber
\\
&&+\ g_a\sigma_z\otimes\tau_z+g_f\sigma_z\otimes1,
\end{eqnarray}
with the basis of $|t\uparrow\rangle$, $|t\downarrow\rangle$, $|b\uparrow\rangle$, and $|b\downarrow\rangle$, where $t$ and $b$ denote the top and bottom surface states, and $\uparrow$ and $\downarrow$ represent spin up and down states, respectively. $\sigma_i$ and $\tau_i$ ($i=x,y,z$) are Pauli matrices acting on spin and layer, respectively. $v_F$ is the Fermi velocity. $m(k)=m_0+m_1(k_x^2+k_y^2)$ is the hybridization between the top and bottom surface states. The third and fourth terms describe the Zeeman-type spin splitting of top $g_t$ and bottom $g_b$ surface states induced by the FM exchange of Mn along $z$ axis, where $g_{a,f}=(g_t\mp g_b)/2$ are the staggered and uniform Zeeman field, respectively. Both $\uparrow$ and $\downarrow$ FM Mn layers will contribute to the Zeeman field. In the mean field approximation, 
\begin{equation}
g_i=\sum_j\text{sgn}(s^z_j)\lambda^{ij},\ \ (i=t,b)
\end{equation}
where $j$ labels the Mn layer index, $s^z_j$ is the $z$-component of Mn local spin in layer $j$, $\lambda^{ij}$ is the effective exchange parameter between local moments in layer $j$ and the top ($i=t$) or bottom ($i=b$) surface states, respectively. All $\lambda^{ij}$ have the same sign and we set $\lambda^{ij}>0$. $\text{sgn}(s^z_j)$ comes from the magnetization direction of each Mn layer.  

The Hamiltonian~(\ref{afm}) describes both QAH and zero plateau states characterized by $\mathcal{C}$. The band dispersion is given by $\varepsilon^2_\pm(k_x,k_y)=v_F^2(k_x^2+k_y^2)+(\sqrt{m(k)^2+g_a^2}\pm g_f)^2$. $\mathcal{C}$ only changes at the gap closing point determined by $\sqrt{m_0^2+g_a^2}=\left|g_f\right|$. When $\sqrt{m_0^2+g_a^2}<\left|g_f\right|$, the system is QAH with $\mathcal{C}=g_f/|g_f|$; while when $\sqrt{m_0^2+g_a^2}>\left|g_f\right|$, the system has $\mathcal{C}=0$. AFM and III are topologically equivalent but have quite different origins. In AFM, with opposite magnetic exchange coupling on two surfaces, $g_f=0$ and $g_a$ is finite, it is magnetic insulator with a nonquantized but finite $\theta$ response~\cite{liuzc2020}. While in III, $g_a=0$ and the hybridization gap exceeds the finite FM exchange gap $g_f$, thus it is a trivial insulator.

With the gap and $\mathcal{C}$ in these magnetic states, $\lambda^{ij}$ can be determined. For $2$ NL, we approximate $\lambda_1\equiv\lambda^{t1}=\lambda^{b4}$, $\lambda_2\equiv\lambda^{t2}=\lambda^{b3}$, $\lambda_3\equiv\lambda^{t3}=\lambda^{b2}$, and $\lambda_4\equiv\lambda^{t4}=\lambda^{b1}$. This yields $\lambda_1=19.0$~meV, $\lambda_2=2.3$~meV, $\lambda_3=18.4$~meV, $\lambda_4=1.8$~meV and $m_0=3.3$~meV. Consistent with $2$ NL decay length of surface states, $\lambda^{ij}$ almost vanishes when $i=t$ and $j\geq5$ as determined from the magnetic states in $3$ NL~\cite{supplementary}.

Finally, we discuss the field-induced magnetic transitions. The evolution of magnetic transition under external field can be described by a magnetic bilayer Stoner-Wohlfarth model with an interlayer exchange coupling $J_{1,2}$ and an effective anisotropy $K$. From Fig.~\ref{fig4}, within each NL the interlayer AFM coupling is $J_1=0.8$~meV, and $J_2=0.35$~meV between adjacent NL. The uniaxial anisotropy $K\approx0.1$~meV in Table~\ref{table}. With the field applied parallel to the magnetic easy $z$ axis, since $K\ll J_{1,2}$, the AFM ground state undergoes a spin-flop transition to a canted state where the sublattice magnetization is roughly perpendicular to $z$ axis. Further increasing the field brings the canted magnetizations to FM state by coherent rotation. The bulk AFM is also energetically favored than bulk III, with energy difference of $0.8$~meV/Mn from first-principles calculation. Take 2 NL as an example, the $z$-axis magnetic field will drive AFM to a canted state at field $H^{\text{sf}}_{1}$, then to II, and finally to FM. For II, it is energetically favorable than I. A rough estimation yields $H^{\text{sf}}_{1}\approx1.6$~T. The coherent rotation of layer magnetization and the corresponding QAH plateau transition (with $\mathcal{C}=0\rightarrow1$) at small fields provides a promising platform for chiral Majorana fermion.

The intrinsic van der Waals magnetic material M$_2$BT hosts rich topological quantum states in different spatial dimensions, which is well described by a Bi$_2$Te$_3$-type TI model with certain 
$\mathcal{T}$-breaking perturbations. We expect superlattice-like new magnetic TI such as M$_2$BT/MBT and M$_2$BT/Bi$_2$Te$_3$ as well as twisted multilayer~\cite{lian2020} with tunable exchange interactions and topological properties may be fabricated. Other tetradymite-type compounds $X_2$Bi$_2$Te$_5$, $X_2$Bi$_2$Se$_5$, and $X_2$Sb$_2$Te$_5$ ($X=$Mn or Eu), if with the same hexagonal crystal structure, are also promising candidates to host magnetic topological states similar to M$_2$BT. This will further enrich the magnetic TI family and provide a new material platform for exotic topological phenomena.

\begin{acknowledgments}
We acknowledge Z. Liu, H. Zhang, W. Li, G. Xu, H. Yang, J. Zhu and Y. Zhang for valuable discussions. This work is supported by the Natural Science Foundation of China through Grant Nos.~11774065, 11574051, and 11874117, the National Key Research Program of China under Grant Nos.~2016YFA0300703 and 2019YFA0308404, Shanghai Municipal Science and Technology Major Project under Grant No.~2019SHZDZX01, and the Natural Science Foundation of Shanghai under Grant No.~19ZR1471400. Y.L. and Y.J. contributed equally to this work.
\end{acknowledgments}


\begin{thebibliography}{73}%
\makeatletter
\providecommand \@ifxundefined [1]{%
 \@ifx{#1\undefined}
}%
\providecommand \@ifnum [1]{%
 \ifnum #1\expandafter \@firstoftwo
 \else \expandafter \@secondoftwo
 \fi
}%
\providecommand \@ifx [1]{%
 \ifx #1\expandafter \@firstoftwo
 \else \expandafter \@secondoftwo
 \fi
}%
\providecommand \natexlab [1]{#1}%
\providecommand \enquote  [1]{``#1''}%
\providecommand \bibnamefont  [1]{#1}%
\providecommand \bibfnamefont [1]{#1}%
\providecommand \citenamefont [1]{#1}%
\providecommand \href@noop [0]{\@secondoftwo}%
\providecommand \href [0]{\begingroup \@sanitize@url \@href}%
\providecommand \@href[1]{\@@startlink{#1}\@@href}%
\providecommand \@@href[1]{\endgroup#1\@@endlink}%
\providecommand \@sanitize@url [0]{\catcode `\\12\catcode `\$12\catcode
  `\&12\catcode `\#12\catcode `\^12\catcode `\_12\catcode `\%12\relax}%
\providecommand \@@startlink[1]{}%
\providecommand \@@endlink[0]{}%
\providecommand \url  [0]{\begingroup\@sanitize@url \@url }%
\providecommand \@url [1]{\endgroup\@href {#1}{\urlprefix }}%
\providecommand \urlprefix  [0]{URL }%
\providecommand \Eprint [0]{\href }%
\providecommand \doibase [0]{http://dx.doi.org/}%
\providecommand \selectlanguage [0]{\@gobble}%
\providecommand \bibinfo  [0]{\@secondoftwo}%
\providecommand \bibfield  [0]{\@secondoftwo}%
\providecommand \translation [1]{[#1]}%
\providecommand \BibitemOpen [0]{}%
\providecommand \bibitemStop [0]{}%
\providecommand \bibitemNoStop [0]{.\EOS\space}%
\providecommand \EOS [0]{\spacefactor3000\relax}%
\providecommand \BibitemShut  [1]{\csname bibitem#1\endcsname}%
\let\auto@bib@innerbib\@empty
\bibitem [{\citenamefont {Kane}\ and\ \citenamefont
  {Mele}(2005{\natexlab{a}})}]{kane2005}%
  \BibitemOpen
  \bibfield  {author} {\bibinfo {author} {\bibfnamefont {C.~L.}\ \bibnamefont
  {Kane}}\ and\ \bibinfo {author} {\bibfnamefont {E.~J.}\ \bibnamefont
  {Mele}},\ }\bibfield  {title} {\enquote {\bibinfo {title} {${Z}_{2}$
  topological order and the quantum spin hall effect},}\ }\href {\doibase
  10.1103/PhysRevLett.95.146802} {\bibfield  {journal} {\bibinfo  {journal}
  {Phys. Rev. Lett.}\ }\textbf {\bibinfo {volume} {95}},\ \bibinfo {pages}
  {146802} (\bibinfo {year} {2005}{\natexlab{a}})}\BibitemShut {NoStop}%
\bibitem [{\citenamefont {Kane}\ and\ \citenamefont
  {Mele}(2005{\natexlab{b}})}]{kane2005b}%
  \BibitemOpen
  \bibfield  {author} {\bibinfo {author} {\bibfnamefont {C.~L.}\ \bibnamefont
  {Kane}}\ and\ \bibinfo {author} {\bibfnamefont {E.~J.}\ \bibnamefont
  {Mele}},\ }\bibfield  {title} {\enquote {\bibinfo {title} {{Quantum Spin Hall
  Effect in Graphene}},}\ }\href {\doibase 10.1103/PhysRevLett.95.226801}
  {\bibfield  {journal} {\bibinfo  {journal} {Phys. Rev. Lett.}\ }\textbf
  {\bibinfo {volume} {95}},\ \bibinfo {pages} {226801} (\bibinfo {year}
  {2005}{\natexlab{b}})}\BibitemShut {NoStop}%
\bibitem [{\citenamefont {Bernevig}\ \emph {et~al.}(2006)\citenamefont
  {Bernevig}, \citenamefont {Hughes},\ and\ \citenamefont
  {Zhang}}]{bernevig2006c}%
  \BibitemOpen
  \bibfield  {author} {\bibinfo {author} {\bibfnamefont {B.~A.}\ \bibnamefont
  {Bernevig}}, \bibinfo {author} {\bibfnamefont {T.~L.}\ \bibnamefont
  {Hughes}}, \ and\ \bibinfo {author} {\bibfnamefont {S.~C.}\ \bibnamefont
  {Zhang}},\ }\bibfield  {title} {\enquote {\bibinfo {title} {\textrm{Quantum
  spin Hall effect and topological phase transition in HgTe quantum wells}},}\
  }\href@noop {} {\bibfield  {journal} {\bibinfo  {journal} {Science}\ }\textbf
  {\bibinfo {volume} {314}},\ \bibinfo {pages} {1757} (\bibinfo {year}
  {2006})}\BibitemShut {NoStop}%
\bibitem [{\citenamefont {K\"onig}\ \emph {et~al.}(2007)\citenamefont
  {K\"onig}, \citenamefont {Wiedmann}, \citenamefont {Br\"une}, \citenamefont
  {Roth}, \citenamefont {Buhmann}, \citenamefont {Molenkamp}, \citenamefont
  {Qi},\ and\ \citenamefont {Zhang}}]{koenig2007}%
  \BibitemOpen
  \bibfield  {author} {\bibinfo {author} {\bibfnamefont {Markus}\ \bibnamefont
  {K\"onig}}, \bibinfo {author} {\bibfnamefont {Steffen}\ \bibnamefont
  {Wiedmann}}, \bibinfo {author} {\bibfnamefont {Christoph}\ \bibnamefont
  {Br\"une}}, \bibinfo {author} {\bibfnamefont {Andreas}\ \bibnamefont {Roth}},
  \bibinfo {author} {\bibfnamefont {Hartmut}\ \bibnamefont {Buhmann}}, \bibinfo
  {author} {\bibfnamefont {Laurens}\ \bibnamefont {Molenkamp}}, \bibinfo
  {author} {\bibfnamefont {Xiao-Liang}\ \bibnamefont {Qi}}, \ and\ \bibinfo
  {author} {\bibfnamefont {Shou-Cheng}\ \bibnamefont {Zhang}},\ }\bibfield
  {title} {\enquote {\bibinfo {title} {{Quantum Spin Hall Insulator State in
  HgTe Quantum Wells}},}\ }\href@noop {} {\bibfield  {journal} {\bibinfo
  {journal} {Science}\ }\textbf {\bibinfo {volume} {318}},\ \bibinfo {pages}
  {766--770} (\bibinfo {year} {2007})}\BibitemShut {NoStop}%
\bibitem [{\citenamefont {Fu}\ \emph {et~al.}(2007)\citenamefont {Fu},
  \citenamefont {Kane},\ and\ \citenamefont {Mele}}]{fu2007}%
  \BibitemOpen
  \bibfield  {author} {\bibinfo {author} {\bibfnamefont {Liang}\ \bibnamefont
  {Fu}}, \bibinfo {author} {\bibfnamefont {C.~L.}\ \bibnamefont {Kane}}, \ and\
  \bibinfo {author} {\bibfnamefont {E.~J.}\ \bibnamefont {Mele}},\ }\bibfield
  {title} {\enquote {\bibinfo {title} {Topological insulators in three
  dimensions},}\ }\href {\doibase 10.1103/PhysRevLett.98.106803} {\bibfield
  {journal} {\bibinfo  {journal} {Phys. Rev. Lett.}\ }\textbf {\bibinfo
  {volume} {98}},\ \bibinfo {pages} {106803} (\bibinfo {year}
  {2007})}\BibitemShut {NoStop}%
\bibitem [{\citenamefont {Hasan}\ and\ \citenamefont {Kane}(2010)}]{hasan2010}%
  \BibitemOpen
  \bibfield  {author} {\bibinfo {author} {\bibfnamefont {M.~Z.}\ \bibnamefont
  {Hasan}}\ and\ \bibinfo {author} {\bibfnamefont {C.~L.}\ \bibnamefont
  {Kane}},\ }\bibfield  {title} {\enquote {\bibinfo {title}
  {\textit{Colloquium}: Topological insulators},}\ }\href {\doibase
  10.1103/RevModPhys.82.3045} {\bibfield  {journal} {\bibinfo  {journal} {Rev.
  Mod. Phys.}\ }\textbf {\bibinfo {volume} {82}},\ \bibinfo {pages}
  {3045--3067} (\bibinfo {year} {2010})}\BibitemShut {NoStop}%
\bibitem [{\citenamefont {Qi}\ and\ \citenamefont {Zhang}(2011)}]{qi2011}%
  \BibitemOpen
  \bibfield  {author} {\bibinfo {author} {\bibfnamefont {Xiao-Liang}\
  \bibnamefont {Qi}}\ and\ \bibinfo {author} {\bibfnamefont {Shou-Cheng}\
  \bibnamefont {Zhang}},\ }\bibfield  {title} {\enquote {\bibinfo {title}
  {Topological insulators and superconductors},}\ }\href {\doibase
  10.1103/RevModPhys.83.1057} {\bibfield  {journal} {\bibinfo  {journal} {Rev.
  Mod. Phys.}\ }\textbf {\bibinfo {volume} {83}},\ \bibinfo {pages}
  {1057--1110} (\bibinfo {year} {2011})}\BibitemShut {NoStop}%
\bibitem [{\citenamefont {Tokura}\ \emph {et~al.}(2019)\citenamefont {Tokura},
  \citenamefont {Yasuda},\ and\ \citenamefont {Tsukazaki}}]{tokura2019}%
  \BibitemOpen
  \bibfield  {author} {\bibinfo {author} {\bibfnamefont {Yoshinori}\
  \bibnamefont {Tokura}}, \bibinfo {author} {\bibfnamefont {Kenji}\
  \bibnamefont {Yasuda}}, \ and\ \bibinfo {author} {\bibfnamefont {Atsushi}\
  \bibnamefont {Tsukazaki}},\ }\bibfield  {title} {\enquote {\bibinfo {title}
  {Magnetic topological insulators},}\ }\href@noop {} {\bibfield  {journal}
  {\bibinfo  {journal} {Nat. Rev. Phys.}\ }\textbf {\bibinfo {volume} {1}},\
  \bibinfo {pages} {126--143} (\bibinfo {year} {2019})}\BibitemShut {NoStop}%
\bibitem [{\citenamefont {Wang}\ and\ \citenamefont {Zhang}(2017)}]{wang2017c}%
  \BibitemOpen
  \bibfield  {author} {\bibinfo {author} {\bibfnamefont {Jing}\ \bibnamefont
  {Wang}}\ and\ \bibinfo {author} {\bibfnamefont {Shou-Cheng}\ \bibnamefont
  {Zhang}},\ }\bibfield  {title} {\enquote {\bibinfo {title} {Topological
  states of condensed matter},}\ }\href@noop {} {\bibfield  {journal} {\bibinfo
   {journal} {Nature Mat.}\ }\textbf {\bibinfo {volume} {16}},\ \bibinfo
  {pages} {1062--1067} (\bibinfo {year} {2017})}\BibitemShut {NoStop}%
\bibitem [{\citenamefont {Haldane}(1988)}]{haldane1988}%
  \BibitemOpen
  \bibfield  {author} {\bibinfo {author} {\bibfnamefont {F.~D.~M.}\
  \bibnamefont {Haldane}},\ }\bibfield  {title} {\enquote {\bibinfo {title}
  {Model for a quantum hall effect without landau levels: Condensed-matter
  realization of the "parity anomaly"},}\ }\href {\doibase
  10.1103/PhysRevLett.61.2015} {\bibfield  {journal} {\bibinfo  {journal}
  {Phys. Rev. Lett.}\ }\textbf {\bibinfo {volume} {61}},\ \bibinfo {pages}
  {2015--2018} (\bibinfo {year} {1988})}\BibitemShut {NoStop}%
\bibitem [{\citenamefont {Qi}\ \emph {et~al.}(2006)\citenamefont {Qi},
  \citenamefont {Wu},\ and\ \citenamefont {Zhang}}]{qi2006}%
  \BibitemOpen
  \bibfield  {author} {\bibinfo {author} {\bibfnamefont {Xiao-Liang}\
  \bibnamefont {Qi}}, \bibinfo {author} {\bibfnamefont {Yong-Shi}\ \bibnamefont
  {Wu}}, \ and\ \bibinfo {author} {\bibfnamefont {Shou-Cheng}\ \bibnamefont
  {Zhang}},\ }\bibfield  {title} {\enquote {\bibinfo {title} {Topological
  quantization of the spin hall effect in two-dimensional paramagnetic
  semiconductors},}\ }\href {\doibase 10.1103/PhysRevB.74.085308} {\bibfield
  {journal} {\bibinfo  {journal} {Phys. Rev. B}\ }\textbf {\bibinfo {volume}
  {74}},\ \bibinfo {pages} {085308} (\bibinfo {year} {2006})}\BibitemShut
  {NoStop}%
\bibitem [{\citenamefont {Liu}\ \emph {et~al.}(2008)\citenamefont {Liu},
  \citenamefont {Qi}, \citenamefont {Dai}, \citenamefont {Fang},\ and\
  \citenamefont {Zhang}}]{liu2008}%
  \BibitemOpen
  \bibfield  {author} {\bibinfo {author} {\bibfnamefont {Chao-Xing}\
  \bibnamefont {Liu}}, \bibinfo {author} {\bibfnamefont {Xiao-Liang}\
  \bibnamefont {Qi}}, \bibinfo {author} {\bibfnamefont {Xi}~\bibnamefont
  {Dai}}, \bibinfo {author} {\bibfnamefont {Zhong}\ \bibnamefont {Fang}}, \
  and\ \bibinfo {author} {\bibfnamefont {Shou-Cheng}\ \bibnamefont {Zhang}},\
  }\bibfield  {title} {\enquote {\bibinfo {title} {Quantum anomalous hall
  effect in $\mathrm{Hg}_{1-y}\mathrm{Mn}_{y}\mathrm{Te}$ quantum wells},}\
  }\href {\doibase 10.1103/PhysRevLett.101.146802} {\bibfield  {journal}
  {\bibinfo  {journal} {Phys. Rev. Lett.}\ }\textbf {\bibinfo {volume} {101}},\
  \bibinfo {pages} {146802} (\bibinfo {year} {2008})}\BibitemShut {NoStop}%
\bibitem [{\citenamefont {Yu}\ \emph {et~al.}(2010)\citenamefont {Yu},
  \citenamefont {Zhang}, \citenamefont {Zhang}, \citenamefont {Zhang},
  \citenamefont {Dai},\ and\ \citenamefont {Fang}}]{yu2010}%
  \BibitemOpen
  \bibfield  {author} {\bibinfo {author} {\bibfnamefont {Rui}\ \bibnamefont
  {Yu}}, \bibinfo {author} {\bibfnamefont {Wei}\ \bibnamefont {Zhang}},
  \bibinfo {author} {\bibfnamefont {Hai-Jun}\ \bibnamefont {Zhang}}, \bibinfo
  {author} {\bibfnamefont {Shou-Cheng}\ \bibnamefont {Zhang}}, \bibinfo
  {author} {\bibfnamefont {Xi}~\bibnamefont {Dai}}, \ and\ \bibinfo {author}
  {\bibfnamefont {Zhong}\ \bibnamefont {Fang}},\ }\bibfield  {title} {\enquote
  {\bibinfo {title} {{Quantized Anomalous Hall Effect in Magnetic Topological
  Insulators}},}\ }\href {\doibase 10.1126/science.1187485} {\bibfield
  {journal} {\bibinfo  {journal} {Science}\ }\textbf {\bibinfo {volume}
  {329}},\ \bibinfo {pages} {61--64} (\bibinfo {year} {2010})}\BibitemShut
  {NoStop}%
\bibitem [{\citenamefont {Wang}\ \emph {et~al.}(2013)\citenamefont {Wang},
  \citenamefont {Lian}, \citenamefont {Zhang}, \citenamefont {Xu},\ and\
  \citenamefont {Zhang}}]{wang2013a}%
  \BibitemOpen
  \bibfield  {author} {\bibinfo {author} {\bibfnamefont {Jing}\ \bibnamefont
  {Wang}}, \bibinfo {author} {\bibfnamefont {Biao}\ \bibnamefont {Lian}},
  \bibinfo {author} {\bibfnamefont {Haijun}\ \bibnamefont {Zhang}}, \bibinfo
  {author} {\bibfnamefont {Yong}\ \bibnamefont {Xu}}, \ and\ \bibinfo {author}
  {\bibfnamefont {Shou-Cheng}\ \bibnamefont {Zhang}},\ }\bibfield  {title}
  {\enquote {\bibinfo {title} {Quantum anomalous hall effect with higher
  plateaus},}\ }\href {\doibase 10.1103/PhysRevLett.111.136801} {\bibfield
  {journal} {\bibinfo  {journal} {Phys. Rev. Lett.}\ }\textbf {\bibinfo
  {volume} {111}},\ \bibinfo {pages} {136801} (\bibinfo {year}
  {2013})}\BibitemShut {NoStop}%
\bibitem [{\citenamefont {Wang}\ \emph
  {et~al.}(2015{\natexlab{a}})\citenamefont {Wang}, \citenamefont {Lian},\ and\
  \citenamefont {Zhang}}]{wang2015d}%
  \BibitemOpen
  \bibfield  {author} {\bibinfo {author} {\bibfnamefont {Jing}\ \bibnamefont
  {Wang}}, \bibinfo {author} {\bibfnamefont {Biao}\ \bibnamefont {Lian}}, \
  and\ \bibinfo {author} {\bibfnamefont {Shou-Cheng}\ \bibnamefont {Zhang}},\
  }\bibfield  {title} {\enquote {\bibinfo {title} {Quantum anomalous hall
  effect in magnetic topological insulators},}\ }\href@noop {} {\bibfield
  {journal} {\bibinfo  {journal} {Phys. Scr.}\ }\textbf {\bibinfo {volume}
  {T164}},\ \bibinfo {pages} {014003} (\bibinfo {year}
  {2015}{\natexlab{a}})}\BibitemShut {NoStop}%
\bibitem [{\citenamefont {Liu}\ \emph {et~al.}(2016)\citenamefont {Liu},
  \citenamefont {Zhang},\ and\ \citenamefont {Qi}}]{liu2016}%
  \BibitemOpen
  \bibfield  {author} {\bibinfo {author} {\bibfnamefont {C.-X.}\ \bibnamefont
  {Liu}}, \bibinfo {author} {\bibfnamefont {S.-C.}\ \bibnamefont {Zhang}}, \
  and\ \bibinfo {author} {\bibfnamefont {X.-L.}\ \bibnamefont {Qi}},\
  }\bibfield  {title} {\enquote {\bibinfo {title} {The quantum anomalous hall
  effect: Theory and experiment},}\ }\href {\doibase
  10.1146/annurev-conmatphys-031115-011417} {\bibfield  {journal} {\bibinfo
  {journal} {Annu. Rev. Condens. Mat. Phys.}\ }\textbf {\bibinfo {volume}
  {7}},\ \bibinfo {pages} {301--321} (\bibinfo {year} {2016})}\BibitemShut
  {NoStop}%
\bibitem [{\citenamefont {Qi}\ \emph {et~al.}(2008)\citenamefont {Qi},
  \citenamefont {Hughes},\ and\ \citenamefont {Zhang}}]{qi2008}%
  \BibitemOpen
  \bibfield  {author} {\bibinfo {author} {\bibfnamefont {Xiao-Liang}\
  \bibnamefont {Qi}}, \bibinfo {author} {\bibfnamefont {Taylor~L.}\
  \bibnamefont {Hughes}}, \ and\ \bibinfo {author} {\bibfnamefont {Shou-Cheng}\
  \bibnamefont {Zhang}},\ }\bibfield  {title} {\enquote {\bibinfo {title}
  {Topological field theory of time-reversal invariant insulators},}\ }\href
  {\doibase 10.1103/PhysRevB.78.195424} {\bibfield  {journal} {\bibinfo
  {journal} {Phys. Rev. B}\ }\textbf {\bibinfo {volume} {78}},\ \bibinfo
  {pages} {195424} (\bibinfo {year} {2008})}\BibitemShut {NoStop}%
\bibitem [{\citenamefont {Essin}\ \emph {et~al.}(2009)\citenamefont {Essin},
  \citenamefont {Moore},\ and\ \citenamefont {Vanderbilt}}]{essin2009}%
  \BibitemOpen
  \bibfield  {author} {\bibinfo {author} {\bibfnamefont {Andrew~M.}\
  \bibnamefont {Essin}}, \bibinfo {author} {\bibfnamefont {Joel~E.}\
  \bibnamefont {Moore}}, \ and\ \bibinfo {author} {\bibfnamefont {David}\
  \bibnamefont {Vanderbilt}},\ }\bibfield  {title} {\enquote {\bibinfo {title}
  {Magnetoelectric polarizability and axion electrodynamics in crystalline
  insulators},}\ }\href {\doibase 10.1103/PhysRevLett.102.146805} {\bibfield
  {journal} {\bibinfo  {journal} {Phys. Rev. Lett.}\ }\textbf {\bibinfo
  {volume} {102}},\ \bibinfo {pages} {146805} (\bibinfo {year}
  {2009})}\BibitemShut {NoStop}%
\bibitem [{\citenamefont {Li}\ \emph {et~al.}(2010)\citenamefont {Li},
  \citenamefont {Wang}, \citenamefont {Qi},\ and\ \citenamefont
  {Zhang}}]{li2010}%
  \BibitemOpen
  \bibfield  {author} {\bibinfo {author} {\bibfnamefont {Rundong}\ \bibnamefont
  {Li}}, \bibinfo {author} {\bibfnamefont {Jing}\ \bibnamefont {Wang}},
  \bibinfo {author} {\bibfnamefont {X.~L.}\ \bibnamefont {Qi}}, \ and\ \bibinfo
  {author} {\bibfnamefont {S.~C.}\ \bibnamefont {Zhang}},\ }\bibfield  {title}
  {\enquote {\bibinfo {title} {Dynamical axion field in topological magnetic
  insulators},}\ }\href@noop {} {\bibfield  {journal} {\bibinfo  {journal}
  {Nature Phys.}\ }\textbf {\bibinfo {volume} {6}},\ \bibinfo {pages} {284}
  (\bibinfo {year} {2010})}\BibitemShut {NoStop}%
\bibitem [{\citenamefont {Wan}\ \emph {et~al.}(2012)\citenamefont {Wan},
  \citenamefont {Vishwanath},\ and\ \citenamefont {Savrasov}}]{wan2012}%
  \BibitemOpen
  \bibfield  {author} {\bibinfo {author} {\bibfnamefont {Xiangang}\
  \bibnamefont {Wan}}, \bibinfo {author} {\bibfnamefont {Ashvin}\ \bibnamefont
  {Vishwanath}}, \ and\ \bibinfo {author} {\bibfnamefont {Sergey~Y.}\
  \bibnamefont {Savrasov}},\ }\bibfield  {title} {\enquote {\bibinfo {title}
  {Computational design of axion insulators based on $5d$ spinel compounds},}\
  }\href {\doibase 10.1103/PhysRevLett.108.146601} {\bibfield  {journal}
  {\bibinfo  {journal} {Phys. Rev. Lett.}\ }\textbf {\bibinfo {volume} {108}},\
  \bibinfo {pages} {146601} (\bibinfo {year} {2012})}\BibitemShut {NoStop}%
\bibitem [{\citenamefont {Wang}\ \emph
  {et~al.}(2015{\natexlab{b}})\citenamefont {Wang}, \citenamefont {Lian},
  \citenamefont {Qi},\ and\ \citenamefont {Zhang}}]{wang2015b}%
  \BibitemOpen
  \bibfield  {author} {\bibinfo {author} {\bibfnamefont {Jing}\ \bibnamefont
  {Wang}}, \bibinfo {author} {\bibfnamefont {Biao}\ \bibnamefont {Lian}},
  \bibinfo {author} {\bibfnamefont {Xiao-Liang}\ \bibnamefont {Qi}}, \ and\
  \bibinfo {author} {\bibfnamefont {Shou-Cheng}\ \bibnamefont {Zhang}},\
  }\bibfield  {title} {\enquote {\bibinfo {title} {{Quantized topological
  magnetoelectric effect of the zero-plateau quantum anomalous Hall state}},}\
  }\href {\doibase 10.1103/PhysRevB.92.081107} {\bibfield  {journal} {\bibinfo
  {journal} {Phys. Rev. B}\ }\textbf {\bibinfo {volume} {92}},\ \bibinfo
  {pages} {081107} (\bibinfo {year} {2015}{\natexlab{b}})}\BibitemShut
  {NoStop}%
\bibitem [{\citenamefont {Morimoto}\ \emph {et~al.}(2015)\citenamefont
  {Morimoto}, \citenamefont {Furusaki},\ and\ \citenamefont
  {Nagaosa}}]{morimoto2015}%
  \BibitemOpen
  \bibfield  {author} {\bibinfo {author} {\bibfnamefont {Takahiro}\
  \bibnamefont {Morimoto}}, \bibinfo {author} {\bibfnamefont {Akira}\
  \bibnamefont {Furusaki}}, \ and\ \bibinfo {author} {\bibfnamefont {Naoto}\
  \bibnamefont {Nagaosa}},\ }\bibfield  {title} {\enquote {\bibinfo {title}
  {Topological magnetoelectric effects in thin films of topological
  insulators},}\ }\href {\doibase 10.1103/PhysRevB.92.085113} {\bibfield
  {journal} {\bibinfo  {journal} {Phys. Rev. B}\ }\textbf {\bibinfo {volume}
  {92}},\ \bibinfo {pages} {085113} (\bibinfo {year} {2015})}\BibitemShut
  {NoStop}%
\bibitem [{\citenamefont {Wang}\ \emph {et~al.}(2016)\citenamefont {Wang},
  \citenamefont {Lian},\ and\ \citenamefont {Zhang}}]{wang2016a}%
  \BibitemOpen
  \bibfield  {author} {\bibinfo {author} {\bibfnamefont {Jing}\ \bibnamefont
  {Wang}}, \bibinfo {author} {\bibfnamefont {Biao}\ \bibnamefont {Lian}}, \
  and\ \bibinfo {author} {\bibfnamefont {Shou-Cheng}\ \bibnamefont {Zhang}},\
  }\bibfield  {title} {\enquote {\bibinfo {title} {Dynamical axion field in a
  magnetic topological insulator superlattice},}\ }\href {\doibase
  10.1103/PhysRevB.93.045115} {\bibfield  {journal} {\bibinfo  {journal} {Phys.
  Rev. B}\ }\textbf {\bibinfo {volume} {93}},\ \bibinfo {pages} {045115}
  (\bibinfo {year} {2016})}\BibitemShut {NoStop}%
\bibitem [{\citenamefont {Mogi}\ \emph {et~al.}(2017)\citenamefont {Mogi},
  \citenamefont {Kawamura}, \citenamefont {Yoshimi}, \citenamefont {Tsukazaki},
  \citenamefont {Kozuka}, \citenamefont {Shirakawa}, \citenamefont {Takahashi},
  \citenamefont {Kawasaki},\ and\ \citenamefont {Tokura}}]{mogi2017}%
  \BibitemOpen
  \bibfield  {author} {\bibinfo {author} {\bibfnamefont {M.}~\bibnamefont
  {Mogi}}, \bibinfo {author} {\bibfnamefont {M.}~\bibnamefont {Kawamura}},
  \bibinfo {author} {\bibfnamefont {R.}~\bibnamefont {Yoshimi}}, \bibinfo
  {author} {\bibfnamefont {A.}~\bibnamefont {Tsukazaki}}, \bibinfo {author}
  {\bibfnamefont {Y.}~\bibnamefont {Kozuka}}, \bibinfo {author} {\bibfnamefont
  {N.}~\bibnamefont {Shirakawa}}, \bibinfo {author} {\bibfnamefont {K.~S.}\
  \bibnamefont {Takahashi}}, \bibinfo {author} {\bibfnamefont {M.}~\bibnamefont
  {Kawasaki}}, \ and\ \bibinfo {author} {\bibfnamefont {Y.}~\bibnamefont
  {Tokura}},\ }\bibfield  {title} {\enquote {\bibinfo {title} {A magnetic
  heterostructure of topological insulators as a candidate for an axion
  insulator},}\ }\href@noop {} {\bibfield  {journal} {\bibinfo  {journal}
  {Nature Mater.}\ }\textbf {\bibinfo {volume} {16}},\ \bibinfo {pages}
  {516--521} (\bibinfo {year} {2017})}\BibitemShut {NoStop}%
\bibitem [{\citenamefont {Xu}\ \emph {et~al.}(2019)\citenamefont {Xu},
  \citenamefont {Song}, \citenamefont {Wang}, \citenamefont {Weng},\ and\
  \citenamefont {Dai}}]{xu2019}%
  \BibitemOpen
  \bibfield  {author} {\bibinfo {author} {\bibfnamefont {Yuanfeng}\
  \bibnamefont {Xu}}, \bibinfo {author} {\bibfnamefont {Zhida}\ \bibnamefont
  {Song}}, \bibinfo {author} {\bibfnamefont {Zhijun}\ \bibnamefont {Wang}},
  \bibinfo {author} {\bibfnamefont {Hongming}\ \bibnamefont {Weng}}, \ and\
  \bibinfo {author} {\bibfnamefont {Xi}~\bibnamefont {Dai}},\ }\bibfield
  {title} {\enquote {\bibinfo {title} {Higher-order topology of the axion
  insulator ${\mathrm{euin}}_{2}{\mathrm{as}}_{2}$},}\ }\href {\doibase
  10.1103/PhysRevLett.122.256402} {\bibfield  {journal} {\bibinfo  {journal}
  {Phys. Rev. Lett.}\ }\textbf {\bibinfo {volume} {122}},\ \bibinfo {pages}
  {256402} (\bibinfo {year} {2019})}\BibitemShut {NoStop}%
\bibitem [{\citenamefont {Chowdhury}\ \emph {et~al.}(2019)\citenamefont
  {Chowdhury}, \citenamefont {Garrity},\ and\ \citenamefont
  {Tavazza}}]{chowdhury2019}%
  \BibitemOpen
  \bibfield  {author} {\bibinfo {author} {\bibfnamefont {Sugata}\ \bibnamefont
  {Chowdhury}}, \bibinfo {author} {\bibfnamefont {Kevin~F.}\ \bibnamefont
  {Garrity}}, \ and\ \bibinfo {author} {\bibfnamefont {Francesca}\ \bibnamefont
  {Tavazza}},\ }\bibfield  {title} {\enquote {\bibinfo {title} {Prediction of
  weyl semimetal and antiferromagnetic topological insulator phases in
  bi2mnse4},}\ }\href@noop {} {\bibfield  {journal} {\bibinfo  {journal} {npj
  Comp. Mat.}\ }\textbf {\bibinfo {volume} {5}},\ \bibinfo {pages} {33}
  (\bibinfo {year} {2019})}\BibitemShut {NoStop}%
\bibitem [{\citenamefont {Read}\ and\ \citenamefont {Green}(2000)}]{read2000}%
  \BibitemOpen
  \bibfield  {author} {\bibinfo {author} {\bibfnamefont {N.}~\bibnamefont
  {Read}}\ and\ \bibinfo {author} {\bibfnamefont {Dmitry}\ \bibnamefont
  {Green}},\ }\bibfield  {title} {\enquote {\bibinfo {title} {Paired states of
  fermions in two dimensions with breaking of parity and time-reversal
  symmetries and the fractional quantum hall effect},}\ }\href {\doibase
  10.1103/PhysRevB.61.10267} {\bibfield  {journal} {\bibinfo  {journal} {Phys.
  Rev. B}\ }\textbf {\bibinfo {volume} {61}},\ \bibinfo {pages} {10267--10297}
  (\bibinfo {year} {2000})}\BibitemShut {NoStop}%
\bibitem [{\citenamefont {Alicea}(2012)}]{alicea2012}%
  \BibitemOpen
  \bibfield  {author} {\bibinfo {author} {\bibfnamefont {Jason}\ \bibnamefont
  {Alicea}},\ }\bibfield  {title} {\enquote {\bibinfo {title} {New directions
  in the pursuit of majorana fermions in solid state systems},}\ }\href@noop {}
  {\bibfield  {journal} {\bibinfo  {journal} {Rep. Prog. Phys.}\ }\textbf
  {\bibinfo {volume} {75}},\ \bibinfo {pages} {076501} (\bibinfo {year}
  {2012})}\BibitemShut {NoStop}%
\bibitem [{\citenamefont {Qi}\ \emph {et~al.}(2010)\citenamefont {Qi},
  \citenamefont {Hughes},\ and\ \citenamefont {Zhang}}]{qi2010b}%
  \BibitemOpen
  \bibfield  {author} {\bibinfo {author} {\bibfnamefont {Xiao-Liang}\
  \bibnamefont {Qi}}, \bibinfo {author} {\bibfnamefont {Taylor~L.}\
  \bibnamefont {Hughes}}, \ and\ \bibinfo {author} {\bibfnamefont {Shou-Cheng}\
  \bibnamefont {Zhang}},\ }\bibfield  {title} {\enquote {\bibinfo {title}
  {Chiral topological superconductor from the quantum hall state},}\ }\href
  {\doibase 10.1103/PhysRevB.82.184516} {\bibfield  {journal} {\bibinfo
  {journal} {Phys. Rev. B}\ }\textbf {\bibinfo {volume} {82}},\ \bibinfo
  {pages} {184516} (\bibinfo {year} {2010})}\BibitemShut {NoStop}%
\bibitem [{\citenamefont {Wang}\ \emph
  {et~al.}(2015{\natexlab{c}})\citenamefont {Wang}, \citenamefont {Zhou},
  \citenamefont {Lian},\ and\ \citenamefont {Zhang}}]{wang2015c}%
  \BibitemOpen
  \bibfield  {author} {\bibinfo {author} {\bibfnamefont {Jing}\ \bibnamefont
  {Wang}}, \bibinfo {author} {\bibfnamefont {Quan}\ \bibnamefont {Zhou}},
  \bibinfo {author} {\bibfnamefont {Biao}\ \bibnamefont {Lian}}, \ and\
  \bibinfo {author} {\bibfnamefont {Shou-Cheng}\ \bibnamefont {Zhang}},\
  }\bibfield  {title} {\enquote {\bibinfo {title} {Chiral topological
  superconductor and half-integer conductance plateau from quantum anomalous
  hall plateau transition},}\ }\href {\doibase 10.1103/PhysRevB.92.064520}
  {\bibfield  {journal} {\bibinfo  {journal} {Phys. Rev. B}\ }\textbf {\bibinfo
  {volume} {92}},\ \bibinfo {pages} {064520} (\bibinfo {year}
  {2015}{\natexlab{c}})}\BibitemShut {NoStop}%
\bibitem [{\citenamefont {Ivanov}(2001)}]{ivanov2001}%
  \BibitemOpen
  \bibfield  {author} {\bibinfo {author} {\bibfnamefont {D.~A.}\ \bibnamefont
  {Ivanov}},\ }\bibfield  {title} {\enquote {\bibinfo {title} {Non-abelian
  statistics of half-quantum vortices in $\mathit{p}$-wave superconductors},}\
  }\href {\doibase 10.1103/PhysRevLett.86.268} {\bibfield  {journal} {\bibinfo
  {journal} {Phys. Rev. Lett.}\ }\textbf {\bibinfo {volume} {86}},\ \bibinfo
  {pages} {268--271} (\bibinfo {year} {2001})}\BibitemShut {NoStop}%
\bibitem [{\citenamefont {Kitaev}(2003)}]{kitaev2003}%
  \BibitemOpen
  \bibfield  {author} {\bibinfo {author} {\bibfnamefont {A.Yu.}\ \bibnamefont
  {Kitaev}},\ }\bibfield  {title} {\enquote {\bibinfo {title} {Fault-tolerant
  quantum computation by anyons},}\ }\href {\doibase
  http://dx.doi.org/10.1016/S0003-4916(02)00018-0} {\bibfield  {journal}
  {\bibinfo  {journal} {Ann. Phys.}\ }\textbf {\bibinfo {volume} {303}},\
  \bibinfo {pages} {2--30} (\bibinfo {year} {2003})}\BibitemShut {NoStop}%
\bibitem [{\citenamefont {Nayak}\ \emph {et~al.}(2008)\citenamefont {Nayak},
  \citenamefont {Simon}, \citenamefont {Stern}, \citenamefont {Freedman},\ and\
  \citenamefont {Das~Sarma}}]{nayak2008}%
  \BibitemOpen
  \bibfield  {author} {\bibinfo {author} {\bibfnamefont {Chetan}\ \bibnamefont
  {Nayak}}, \bibinfo {author} {\bibfnamefont {Steven~H.}\ \bibnamefont
  {Simon}}, \bibinfo {author} {\bibfnamefont {Ady}\ \bibnamefont {Stern}},
  \bibinfo {author} {\bibfnamefont {Michael}\ \bibnamefont {Freedman}}, \ and\
  \bibinfo {author} {\bibfnamefont {Sankar}\ \bibnamefont {Das~Sarma}},\
  }\bibfield  {title} {\enquote {\bibinfo {title} {Non-abelian anyons and
  topological quantum computation},}\ }\href {\doibase
  10.1103/RevModPhys.80.1083} {\bibfield  {journal} {\bibinfo  {journal} {Rev.
  Mod. Phys.}\ }\textbf {\bibinfo {volume} {80}},\ \bibinfo {pages}
  {1083--1159} (\bibinfo {year} {2008})}\BibitemShut {NoStop}%
\bibitem [{\citenamefont {Alicea}\ \emph {et~al.}(2011)\citenamefont {Alicea},
  \citenamefont {Oreg}, \citenamefont {Refael}, \citenamefont {von Oppen},\
  and\ \citenamefont {Fisher}}]{alicea2011}%
  \BibitemOpen
  \bibfield  {author} {\bibinfo {author} {\bibfnamefont {Jason}\ \bibnamefont
  {Alicea}}, \bibinfo {author} {\bibfnamefont {Yuval}\ \bibnamefont {Oreg}},
  \bibinfo {author} {\bibfnamefont {Gil}\ \bibnamefont {Refael}}, \bibinfo
  {author} {\bibfnamefont {Felix}\ \bibnamefont {von Oppen}}, \ and\ \bibinfo
  {author} {\bibfnamefont {Matthew P.~A.}\ \bibnamefont {Fisher}},\ }\bibfield
  {title} {\enquote {\bibinfo {title} {Non-abelian statistics and topological
  quantum information processing in 1d wire networks},}\ }\href@noop {}
  {\bibfield  {journal} {\bibinfo  {journal} {Nature Phys.}\ }\textbf {\bibinfo
  {volume} {7}},\ \bibinfo {pages} {412--417} (\bibinfo {year}
  {2011})}\BibitemShut {NoStop}%
\bibitem [{\citenamefont {Lian}\ \emph
  {et~al.}(2018{\natexlab{a}})\citenamefont {Lian}, \citenamefont {Sun},
  \citenamefont {Vaezi}, \citenamefont {Qi},\ and\ \citenamefont
  {Zhang}}]{lian2018b}%
  \BibitemOpen
  \bibfield  {author} {\bibinfo {author} {\bibfnamefont {B.}~\bibnamefont
  {Lian}}, \bibinfo {author} {\bibfnamefont {X.-Q.}\ \bibnamefont {Sun}},
  \bibinfo {author} {\bibfnamefont {A.}~\bibnamefont {Vaezi}}, \bibinfo
  {author} {\bibfnamefont {X.-L.}\ \bibnamefont {Qi}}, \ and\ \bibinfo {author}
  {\bibfnamefont {S.-C.}\ \bibnamefont {Zhang}},\ }\bibfield  {title} {\enquote
  {\bibinfo {title} {Topological quantum computation based on chiral majorana
  fermions},}\ }\href {\doibase 10.1073/pnas.1810003115} {\bibfield  {journal}
  {\bibinfo  {journal} {Proc. Natl. Acad. Sci. U.S.A.}\ }\textbf {\bibinfo
  {volume} {115}},\ \bibinfo {pages} {10938--10942} (\bibinfo {year}
  {2018}{\natexlab{a}})}\BibitemShut {NoStop}%
\bibitem [{\citenamefont {Chang}\ \emph {et~al.}(2013)\citenamefont {Chang},
  \citenamefont {Zhang}, \citenamefont {Feng}, \citenamefont {Shen},
  \citenamefont {Zhang}, \citenamefont {Guo}, \citenamefont {Li}, \citenamefont
  {Ou}, \citenamefont {Wei}, \citenamefont {Wang}, \citenamefont {Ji},
  \citenamefont {Feng}, \citenamefont {Ji}, \citenamefont {Chen}, \citenamefont
  {Jia}, \citenamefont {Dai}, \citenamefont {Fang}, \citenamefont {Zhang},
  \citenamefont {He}, \citenamefont {Wang}, \citenamefont {Lu}, \citenamefont
  {Ma},\ and\ \citenamefont {Xue}}]{chang2013b}%
  \BibitemOpen
  \bibfield  {author} {\bibinfo {author} {\bibfnamefont {Cui-Zu}\ \bibnamefont
  {Chang}}, \bibinfo {author} {\bibfnamefont {Jinsong}\ \bibnamefont {Zhang}},
  \bibinfo {author} {\bibfnamefont {Xiao}\ \bibnamefont {Feng}}, \bibinfo
  {author} {\bibfnamefont {Jie}\ \bibnamefont {Shen}}, \bibinfo {author}
  {\bibfnamefont {Zuocheng}\ \bibnamefont {Zhang}}, \bibinfo {author}
  {\bibfnamefont {Minghua}\ \bibnamefont {Guo}}, \bibinfo {author}
  {\bibfnamefont {Kang}\ \bibnamefont {Li}}, \bibinfo {author} {\bibfnamefont
  {Yunbo}\ \bibnamefont {Ou}}, \bibinfo {author} {\bibfnamefont {Pang}\
  \bibnamefont {Wei}}, \bibinfo {author} {\bibfnamefont {Li-Li}\ \bibnamefont
  {Wang}}, \bibinfo {author} {\bibfnamefont {Zhong-Qing}\ \bibnamefont {Ji}},
  \bibinfo {author} {\bibfnamefont {Yang}\ \bibnamefont {Feng}}, \bibinfo
  {author} {\bibfnamefont {Shuaihua}\ \bibnamefont {Ji}}, \bibinfo {author}
  {\bibfnamefont {Xi}~\bibnamefont {Chen}}, \bibinfo {author} {\bibfnamefont
  {Jinfeng}\ \bibnamefont {Jia}}, \bibinfo {author} {\bibfnamefont
  {Xi}~\bibnamefont {Dai}}, \bibinfo {author} {\bibfnamefont {Zhong}\
  \bibnamefont {Fang}}, \bibinfo {author} {\bibfnamefont {Shou-Cheng}\
  \bibnamefont {Zhang}}, \bibinfo {author} {\bibfnamefont {Ke}~\bibnamefont
  {He}}, \bibinfo {author} {\bibfnamefont {Yayu}\ \bibnamefont {Wang}},
  \bibinfo {author} {\bibfnamefont {Li}~\bibnamefont {Lu}}, \bibinfo {author}
  {\bibfnamefont {Xu-Cun}\ \bibnamefont {Ma}}, \ and\ \bibinfo {author}
  {\bibfnamefont {Qi-Kun}\ \bibnamefont {Xue}},\ }\bibfield  {title} {\enquote
  {\bibinfo {title} {{Experimental Observation of the Quantum Anomalous Hall
  Effect in a Magnetic Topological Insulator}},}\ }\href {\doibase
  10.1126/science.1234414} {\bibfield  {journal} {\bibinfo  {journal}
  {Science}\ }\textbf {\bibinfo {volume} {340}},\ \bibinfo {pages} {167--170}
  (\bibinfo {year} {2013})}\BibitemShut {NoStop}%
\bibitem [{\citenamefont {Checkelsky}\ \emph {et~al.}(2014)\citenamefont
  {Checkelsky}, \citenamefont {Yoshimi}, \citenamefont {Tsukazaki},
  \citenamefont {Takahashi}, \citenamefont {Kozuka}, \citenamefont {Falson},
  \citenamefont {Kawasaki},\ and\ \citenamefont {Tokura}}]{checkelsky2014}%
  \BibitemOpen
  \bibfield  {author} {\bibinfo {author} {\bibfnamefont {J.~G.}\ \bibnamefont
  {Checkelsky}}, \bibinfo {author} {\bibfnamefont {R.}~\bibnamefont {Yoshimi}},
  \bibinfo {author} {\bibfnamefont {A.}~\bibnamefont {Tsukazaki}}, \bibinfo
  {author} {\bibfnamefont {K.~S.}\ \bibnamefont {Takahashi}}, \bibinfo {author}
  {\bibfnamefont {Y.}~\bibnamefont {Kozuka}}, \bibinfo {author} {\bibfnamefont
  {J.}~\bibnamefont {Falson}}, \bibinfo {author} {\bibfnamefont
  {M.}~\bibnamefont {Kawasaki}}, \ and\ \bibinfo {author} {\bibfnamefont
  {Y.}~\bibnamefont {Tokura}},\ }\bibfield  {title} {\enquote {\bibinfo {title}
  {Trajectory of the anomalous hall effect towards the quantized state in a
  ferromagnetic topological insulator},}\ }\href@noop {} {\bibfield  {journal}
  {\bibinfo  {journal} {Nature Phys.}\ }\textbf {\bibinfo {volume} {10}},\
  \bibinfo {pages} {731} (\bibinfo {year} {2014})}\BibitemShut {NoStop}%
\bibitem [{\citenamefont {Bestwick}\ \emph {et~al.}(2015)\citenamefont
  {Bestwick}, \citenamefont {Fox}, \citenamefont {Kou}, \citenamefont {Pan},
  \citenamefont {Wang},\ and\ \citenamefont {Goldhaber-Gordon}}]{bestwick2015}%
  \BibitemOpen
  \bibfield  {author} {\bibinfo {author} {\bibfnamefont {A.~J.}\ \bibnamefont
  {Bestwick}}, \bibinfo {author} {\bibfnamefont {E.~J.}\ \bibnamefont {Fox}},
  \bibinfo {author} {\bibfnamefont {Xufeng}\ \bibnamefont {Kou}}, \bibinfo
  {author} {\bibfnamefont {Lei}\ \bibnamefont {Pan}}, \bibinfo {author}
  {\bibfnamefont {Kang~L.}\ \bibnamefont {Wang}}, \ and\ \bibinfo {author}
  {\bibfnamefont {D.}~\bibnamefont {Goldhaber-Gordon}},\ }\bibfield  {title}
  {\enquote {\bibinfo {title} {Precise quantization of the anomalous hall
  effect near zero magnetic field},}\ }\href {\doibase
  10.1103/PhysRevLett.114.187201} {\bibfield  {journal} {\bibinfo  {journal}
  {Phys. Rev. Lett.}\ }\textbf {\bibinfo {volume} {114}},\ \bibinfo {pages}
  {187201} (\bibinfo {year} {2015})}\BibitemShut {NoStop}%
\bibitem [{\citenamefont {Chang}\ \emph {et~al.}(2015)\citenamefont {Chang},
  \citenamefont {Zhao}, \citenamefont {Kim}, \citenamefont {Zhang},
  \citenamefont {Assaf}, \citenamefont {Heiman}, \citenamefont {Zhang},
  \citenamefont {Liu}, \citenamefont {Chan},\ and\ \citenamefont
  {Moodera}}]{chang2015}%
  \BibitemOpen
  \bibfield  {author} {\bibinfo {author} {\bibfnamefont {Cui-Zu}\ \bibnamefont
  {Chang}}, \bibinfo {author} {\bibfnamefont {Weiwei}\ \bibnamefont {Zhao}},
  \bibinfo {author} {\bibfnamefont {Duk~Y.}\ \bibnamefont {Kim}}, \bibinfo
  {author} {\bibfnamefont {Haijun}\ \bibnamefont {Zhang}}, \bibinfo {author}
  {\bibfnamefont {Badih~A.}\ \bibnamefont {Assaf}}, \bibinfo {author}
  {\bibfnamefont {Don}\ \bibnamefont {Heiman}}, \bibinfo {author}
  {\bibfnamefont {Shou-Cheng}\ \bibnamefont {Zhang}}, \bibinfo {author}
  {\bibfnamefont {Chaoxing}\ \bibnamefont {Liu}}, \bibinfo {author}
  {\bibfnamefont {Moses H.~W.}\ \bibnamefont {Chan}}, \ and\ \bibinfo {author}
  {\bibfnamefont {Jagadeesh~S.}\ \bibnamefont {Moodera}},\ }\bibfield  {title}
  {\enquote {\bibinfo {title} {High-precision realization of robust quantum
  anomalous hall state in a hard ferromagnetic topological insulator},}\
  }\href@noop {} {\bibfield  {journal} {\bibinfo  {journal} {Nature Mater.}\
  }\textbf {\bibinfo {volume} {14}},\ \bibinfo {pages} {473} (\bibinfo {year}
  {2015})}\BibitemShut {NoStop}%
\bibitem [{\citenamefont {Lee}\ \emph {et~al.}(2015)\citenamefont {Lee},
  \citenamefont {Kim}, \citenamefont {Lee}, \citenamefont {Billinge},
  \citenamefont {Zhong}, \citenamefont {Schneeloch}, \citenamefont {Liu},
  \citenamefont {Valla}, \citenamefont {Tranquada}, \citenamefont {Gu},\ and\
  \citenamefont {Davis}}]{lee2015}%
  \BibitemOpen
  \bibfield  {author} {\bibinfo {author} {\bibfnamefont {Inhee}\ \bibnamefont
  {Lee}}, \bibinfo {author} {\bibfnamefont {Chung~Koo}\ \bibnamefont {Kim}},
  \bibinfo {author} {\bibfnamefont {Jinho}\ \bibnamefont {Lee}}, \bibinfo
  {author} {\bibfnamefont {Simon J.~L.}\ \bibnamefont {Billinge}}, \bibinfo
  {author} {\bibfnamefont {Ruidan}\ \bibnamefont {Zhong}}, \bibinfo {author}
  {\bibfnamefont {John~A.}\ \bibnamefont {Schneeloch}}, \bibinfo {author}
  {\bibfnamefont {Tiansheng}\ \bibnamefont {Liu}}, \bibinfo {author}
  {\bibfnamefont {Tonica}\ \bibnamefont {Valla}}, \bibinfo {author}
  {\bibfnamefont {John~M.}\ \bibnamefont {Tranquada}}, \bibinfo {author}
  {\bibfnamefont {Genda}\ \bibnamefont {Gu}}, \ and\ \bibinfo {author}
  {\bibfnamefont {J.~C.~S{\'e}amus}\ \bibnamefont {Davis}},\ }\bibfield
  {title} {\enquote {\bibinfo {title} {Imaging dirac-mass disorder from
  magnetic dopant atoms in the ferromagnetic topological insulator
  crx(bi0.1sb0.9)2-xte3},}\ }\href {\doibase 10.1073/pnas.1424322112}
  {\bibfield  {journal} {\bibinfo  {journal} {Proc. Natl. Acad. Sci. USA}\
  }\textbf {\bibinfo {volume} {112}},\ \bibinfo {pages} {1316--1321} (\bibinfo
  {year} {2015})}\BibitemShut {NoStop}%
\bibitem [{\citenamefont {He}\ \emph {et~al.}(2017)\citenamefont {He},
  \citenamefont {Pan}, \citenamefont {Stern}, \citenamefont {Burks},
  \citenamefont {Che}, \citenamefont {Yin}, \citenamefont {Wang}, \citenamefont
  {Lian}, \citenamefont {Zhou}, \citenamefont {Choi}, \citenamefont {Murata},
  \citenamefont {Kou}, \citenamefont {Chen}, \citenamefont {Nie}, \citenamefont
  {Shao}, \citenamefont {Fan}, \citenamefont {Zhang}, \citenamefont {Liu},
  \citenamefont {Xia},\ and\ \citenamefont {Wang}}]{he2017}%
  \BibitemOpen
  \bibfield  {author} {\bibinfo {author} {\bibfnamefont {Qing~Lin}\
  \bibnamefont {He}}, \bibinfo {author} {\bibfnamefont {Lei}\ \bibnamefont
  {Pan}}, \bibinfo {author} {\bibfnamefont {Alexander~L.}\ \bibnamefont
  {Stern}}, \bibinfo {author} {\bibfnamefont {Edward~C.}\ \bibnamefont
  {Burks}}, \bibinfo {author} {\bibfnamefont {Xiaoyu}\ \bibnamefont {Che}},
  \bibinfo {author} {\bibfnamefont {Gen}\ \bibnamefont {Yin}}, \bibinfo
  {author} {\bibfnamefont {Jing}\ \bibnamefont {Wang}}, \bibinfo {author}
  {\bibfnamefont {Biao}\ \bibnamefont {Lian}}, \bibinfo {author} {\bibfnamefont
  {Quan}\ \bibnamefont {Zhou}}, \bibinfo {author} {\bibfnamefont {Eun~Sang}\
  \bibnamefont {Choi}}, \bibinfo {author} {\bibfnamefont {Koichi}\ \bibnamefont
  {Murata}}, \bibinfo {author} {\bibfnamefont {Xufeng}\ \bibnamefont {Kou}},
  \bibinfo {author} {\bibfnamefont {Zhijie}\ \bibnamefont {Chen}}, \bibinfo
  {author} {\bibfnamefont {Tianxiao}\ \bibnamefont {Nie}}, \bibinfo {author}
  {\bibfnamefont {Qiming}\ \bibnamefont {Shao}}, \bibinfo {author}
  {\bibfnamefont {Yabin}\ \bibnamefont {Fan}}, \bibinfo {author} {\bibfnamefont
  {Shou-Cheng}\ \bibnamefont {Zhang}}, \bibinfo {author} {\bibfnamefont {Kai}\
  \bibnamefont {Liu}}, \bibinfo {author} {\bibfnamefont {Jing}\ \bibnamefont
  {Xia}}, \ and\ \bibinfo {author} {\bibfnamefont {Kang~L.}\ \bibnamefont
  {Wang}},\ }\bibfield  {title} {\enquote {\bibinfo {title} {Chiral majorana
  fermion modes in a quantum anomalous hall
  insulator{\textendash}superconductor structure},}\ }\href {\doibase
  10.1126/science.aag2792} {\bibfield  {journal} {\bibinfo  {journal}
  {Science}\ }\textbf {\bibinfo {volume} {357}},\ \bibinfo {pages} {294--299}
  (\bibinfo {year} {2017})}\BibitemShut {NoStop}%
\bibitem [{\citenamefont {Kayyalha}\ \emph {et~al.}(2020)\citenamefont
  {Kayyalha}, \citenamefont {Xiao}, \citenamefont {Zhang}, \citenamefont
  {Shin}, \citenamefont {Jiang}, \citenamefont {Wang}, \citenamefont {Zhao},
  \citenamefont {Xiao}, \citenamefont {Zhang}, \citenamefont {Fijalkowski},
  \citenamefont {Mandal}, \citenamefont {Winnerlein}, \citenamefont {Gould},
  \citenamefont {Li}, \citenamefont {Molenkamp}, \citenamefont {Chan},
  \citenamefont {Samarth},\ and\ \citenamefont {Chang}}]{kayyalha2020}%
  \BibitemOpen
  \bibfield  {author} {\bibinfo {author} {\bibfnamefont {Morteza}\ \bibnamefont
  {Kayyalha}}, \bibinfo {author} {\bibfnamefont {Di}~\bibnamefont {Xiao}},
  \bibinfo {author} {\bibfnamefont {Ruoxi}\ \bibnamefont {Zhang}}, \bibinfo
  {author} {\bibfnamefont {Jaeho}\ \bibnamefont {Shin}}, \bibinfo {author}
  {\bibfnamefont {Jue}\ \bibnamefont {Jiang}}, \bibinfo {author} {\bibfnamefont
  {Fei}\ \bibnamefont {Wang}}, \bibinfo {author} {\bibfnamefont {Yi-Fan}\
  \bibnamefont {Zhao}}, \bibinfo {author} {\bibfnamefont {Run}\ \bibnamefont
  {Xiao}}, \bibinfo {author} {\bibfnamefont {Ling}\ \bibnamefont {Zhang}},
  \bibinfo {author} {\bibfnamefont {Kajetan~M.}\ \bibnamefont {Fijalkowski}},
  \bibinfo {author} {\bibfnamefont {Pankaj}\ \bibnamefont {Mandal}}, \bibinfo
  {author} {\bibfnamefont {Martin}\ \bibnamefont {Winnerlein}}, \bibinfo
  {author} {\bibfnamefont {Charles}\ \bibnamefont {Gould}}, \bibinfo {author}
  {\bibfnamefont {Qi}~\bibnamefont {Li}}, \bibinfo {author} {\bibfnamefont
  {Laurens~W.}\ \bibnamefont {Molenkamp}}, \bibinfo {author} {\bibfnamefont
  {Moses H.~W.}\ \bibnamefont {Chan}}, \bibinfo {author} {\bibfnamefont
  {Nitin}\ \bibnamefont {Samarth}}, \ and\ \bibinfo {author} {\bibfnamefont
  {Cui-Zu}\ \bibnamefont {Chang}},\ }\bibfield  {title} {\enquote {\bibinfo
  {title} {Absence of evidence for chiral majorana modes in quantum anomalous
  hall-superconductor devices},}\ }\href {\doibase 10.1126/science.aax6361}
  {\bibfield  {journal} {\bibinfo  {journal} {Science}\ }\textbf {\bibinfo
  {volume} {367}},\ \bibinfo {pages} {64--67} (\bibinfo {year}
  {2020})}\BibitemShut {NoStop}%
\bibitem [{\citenamefont {Ji}\ and\ \citenamefont {Wen}(2018)}]{ji2018}%
  \BibitemOpen
  \bibfield  {author} {\bibinfo {author} {\bibfnamefont {Wenjie}\ \bibnamefont
  {Ji}}\ and\ \bibinfo {author} {\bibfnamefont {Xiao-Gang}\ \bibnamefont
  {Wen}},\ }\bibfield  {title} {\enquote {\bibinfo {title}
  {$\frac{1}{2}({e}^{2}/h)$ conductance plateau without 1d chiral majorana
  fermions},}\ }\href {\doibase 10.1103/PhysRevLett.120.107002} {\bibfield
  {journal} {\bibinfo  {journal} {Phys. Rev. Lett.}\ }\textbf {\bibinfo
  {volume} {120}},\ \bibinfo {pages} {107002} (\bibinfo {year}
  {2018})}\BibitemShut {NoStop}%
\bibitem [{\citenamefont {Huang}\ \emph {et~al.}(2018)\citenamefont {Huang},
  \citenamefont {Setiawan},\ and\ \citenamefont {Sau}}]{huang2018}%
  \BibitemOpen
  \bibfield  {author} {\bibinfo {author} {\bibfnamefont {Yingyi}\ \bibnamefont
  {Huang}}, \bibinfo {author} {\bibfnamefont {F.}~\bibnamefont {Setiawan}}, \
  and\ \bibinfo {author} {\bibfnamefont {Jay~D.}\ \bibnamefont {Sau}},\
  }\bibfield  {title} {\enquote {\bibinfo {title} {Disorder-induced
  half-integer quantized conductance plateau in quantum anomalous hall
  insulator-superconductor structures},}\ }\href {\doibase
  10.1103/PhysRevB.97.100501} {\bibfield  {journal} {\bibinfo  {journal} {Phys.
  Rev. B}\ }\textbf {\bibinfo {volume} {97}},\ \bibinfo {pages} {100501}
  (\bibinfo {year} {2018})}\BibitemShut {NoStop}%
\bibitem [{\citenamefont {Lian}\ \emph
  {et~al.}(2018{\natexlab{b}})\citenamefont {Lian}, \citenamefont {Wang},
  \citenamefont {Sun}, \citenamefont {Vaezi},\ and\ \citenamefont
  {Zhang}}]{lian2018}%
  \BibitemOpen
  \bibfield  {author} {\bibinfo {author} {\bibfnamefont {Biao}\ \bibnamefont
  {Lian}}, \bibinfo {author} {\bibfnamefont {Jing}\ \bibnamefont {Wang}},
  \bibinfo {author} {\bibfnamefont {Xiao-Qi}\ \bibnamefont {Sun}}, \bibinfo
  {author} {\bibfnamefont {Abolhassan}\ \bibnamefont {Vaezi}}, \ and\ \bibinfo
  {author} {\bibfnamefont {Shou-Cheng}\ \bibnamefont {Zhang}},\ }\bibfield
  {title} {\enquote {\bibinfo {title} {Quantum phase transition of chiral
  majorana fermions in the presence of disorder},}\ }\href {\doibase
  10.1103/PhysRevB.97.125408} {\bibfield  {journal} {\bibinfo  {journal} {Phys.
  Rev. B}\ }\textbf {\bibinfo {volume} {97}},\ \bibinfo {pages} {125408}
  (\bibinfo {year} {2018}{\natexlab{b}})}\BibitemShut {NoStop}%
\bibitem [{\citenamefont {Zhang}\ \emph {et~al.}(2019)\citenamefont {Zhang},
  \citenamefont {Shi}, \citenamefont {Zhu}, \citenamefont {Xing}, \citenamefont
  {Zhang},\ and\ \citenamefont {Wang}}]{zhang2019}%
  \BibitemOpen
  \bibfield  {author} {\bibinfo {author} {\bibfnamefont {Dongqin}\ \bibnamefont
  {Zhang}}, \bibinfo {author} {\bibfnamefont {Minji}\ \bibnamefont {Shi}},
  \bibinfo {author} {\bibfnamefont {Tongshuai}\ \bibnamefont {Zhu}}, \bibinfo
  {author} {\bibfnamefont {Dingyu}\ \bibnamefont {Xing}}, \bibinfo {author}
  {\bibfnamefont {Haijun}\ \bibnamefont {Zhang}}, \ and\ \bibinfo {author}
  {\bibfnamefont {Jing}\ \bibnamefont {Wang}},\ }\bibfield  {title} {\enquote
  {\bibinfo {title} {Topological axion states in the magnetic insulator
  ${\mathrm{mnbi}}_{2}{\mathrm{te}}_{4}$ with the quantized magnetoelectric
  effect},}\ }\href {\doibase 10.1103/PhysRevLett.122.206401} {\bibfield
  {journal} {\bibinfo  {journal} {Phys. Rev. Lett.}\ }\textbf {\bibinfo
  {volume} {122}},\ \bibinfo {pages} {206401} (\bibinfo {year}
  {2019})}\BibitemShut {NoStop}%
\bibitem [{\citenamefont {Li}\ \emph {et~al.}(2019{\natexlab{a}})\citenamefont
  {Li}, \citenamefont {Li}, \citenamefont {Du}, \citenamefont {Wang},
  \citenamefont {Gu}, \citenamefont {Zhang}, \citenamefont {He}, \citenamefont
  {Duan},\ and\ \citenamefont {Xu}}]{li2019}%
  \BibitemOpen
  \bibfield  {author} {\bibinfo {author} {\bibfnamefont {Jiaheng}\ \bibnamefont
  {Li}}, \bibinfo {author} {\bibfnamefont {Yang}\ \bibnamefont {Li}}, \bibinfo
  {author} {\bibfnamefont {Shiqiao}\ \bibnamefont {Du}}, \bibinfo {author}
  {\bibfnamefont {Zun}\ \bibnamefont {Wang}}, \bibinfo {author} {\bibfnamefont
  {Bing-Lin}\ \bibnamefont {Gu}}, \bibinfo {author} {\bibfnamefont
  {Shou-Cheng}\ \bibnamefont {Zhang}}, \bibinfo {author} {\bibfnamefont
  {Ke}~\bibnamefont {He}}, \bibinfo {author} {\bibfnamefont {Wenhui}\
  \bibnamefont {Duan}}, \ and\ \bibinfo {author} {\bibfnamefont {Yong}\
  \bibnamefont {Xu}},\ }\bibfield  {title} {\enquote {\bibinfo {title}
  {Intrinsic magnetic topological insulators in van der waals layered
  mnbi2te4-family materials},}\ }\href {\doibase 10.1126/sciadv.aaw5685}
  {\bibfield  {journal} {\bibinfo  {journal} {Sci. Adv.}\ }\textbf {\bibinfo
  {volume} {5}},\ \bibinfo {pages} {eaaw5685} (\bibinfo {year}
  {2019}{\natexlab{a}})}\BibitemShut {NoStop}%
\bibitem [{\citenamefont {Gong}\ \emph {et~al.}(2019)\citenamefont {Gong},
  \citenamefont {Guo}, \citenamefont {Li}, \citenamefont {Zhu}, \citenamefont
  {Liao}, \citenamefont {Liu}, \citenamefont {Zhang}, \citenamefont {Gu},
  \citenamefont {Tang}, \citenamefont {Feng}, \citenamefont {Zhang},
  \citenamefont {Li}, \citenamefont {Song}, \citenamefont {Wang}, \citenamefont
  {Yu}, \citenamefont {Chen}, \citenamefont {Wang}, \citenamefont {Yao},
  \citenamefont {Duan}, \citenamefont {Xu}, \citenamefont {Zhang},
  \citenamefont {Ma}, \citenamefont {Xue},\ and\ \citenamefont
  {He}}]{gong2019}%
  \BibitemOpen
  \bibfield  {author} {\bibinfo {author} {\bibfnamefont {Yan}\ \bibnamefont
  {Gong}}, \bibinfo {author} {\bibfnamefont {Jingwen}\ \bibnamefont {Guo}},
  \bibinfo {author} {\bibfnamefont {Jiaheng}\ \bibnamefont {Li}}, \bibinfo
  {author} {\bibfnamefont {Kejing}\ \bibnamefont {Zhu}}, \bibinfo {author}
  {\bibfnamefont {Menghan}\ \bibnamefont {Liao}}, \bibinfo {author}
  {\bibfnamefont {Xiaozhi}\ \bibnamefont {Liu}}, \bibinfo {author}
  {\bibfnamefont {Qinghua}\ \bibnamefont {Zhang}}, \bibinfo {author}
  {\bibfnamefont {Lin}\ \bibnamefont {Gu}}, \bibinfo {author} {\bibfnamefont
  {Lin}\ \bibnamefont {Tang}}, \bibinfo {author} {\bibfnamefont {Xiao}\
  \bibnamefont {Feng}}, \bibinfo {author} {\bibfnamefont {Ding}\ \bibnamefont
  {Zhang}}, \bibinfo {author} {\bibfnamefont {Wei}\ \bibnamefont {Li}},
  \bibinfo {author} {\bibfnamefont {Canli}\ \bibnamefont {Song}}, \bibinfo
  {author} {\bibfnamefont {Lili}\ \bibnamefont {Wang}}, \bibinfo {author}
  {\bibfnamefont {Pu}~\bibnamefont {Yu}}, \bibinfo {author} {\bibfnamefont
  {Xi}~\bibnamefont {Chen}}, \bibinfo {author} {\bibfnamefont {Yayu}\
  \bibnamefont {Wang}}, \bibinfo {author} {\bibfnamefont {Hong}\ \bibnamefont
  {Yao}}, \bibinfo {author} {\bibfnamefont {Wenhui}\ \bibnamefont {Duan}},
  \bibinfo {author} {\bibfnamefont {Yong}\ \bibnamefont {Xu}}, \bibinfo
  {author} {\bibfnamefont {Shou-Cheng}\ \bibnamefont {Zhang}}, \bibinfo
  {author} {\bibfnamefont {Xucun}\ \bibnamefont {Ma}}, \bibinfo {author}
  {\bibfnamefont {Qi-Kun}\ \bibnamefont {Xue}}, \ and\ \bibinfo {author}
  {\bibfnamefont {Ke}~\bibnamefont {He}},\ }\bibfield  {title} {\enquote
  {\bibinfo {title} {Experimental realization of an intrinsic magnetic
  topological insulator},}\ }\href {\doibase 10.1088/0256-307X/36/7/076801}
  {\bibfield  {journal} {\bibinfo  {journal} {Chin. Phys. Lett.}\ }\textbf
  {\bibinfo {volume} {36}},\ \bibinfo {eid} {076801} (\bibinfo {year}
  {2019})}\BibitemShut {NoStop}%
\bibitem [{\citenamefont {{Otrokov}}\ \emph {et~al.}(2019)\citenamefont
  {{Otrokov}}, \citenamefont {{Klimovskikh}}, \citenamefont {{Bentmann}},
  \citenamefont {{Zeugner}}, \citenamefont {{Aliev}}, \citenamefont {{Gass}},
  \citenamefont {{Wolter}}, \citenamefont {{Koroleva}}, \citenamefont
  {{Estyunin}}, \citenamefont {{Shikin}}, \citenamefont {{Blanco-Rey}},
  \citenamefont {{Hoffmann}}, \citenamefont {{Vyazovskaya}}, \citenamefont
  {{Eremeev}}, \citenamefont {{Koroteev}}, \citenamefont {{Amiraslanov}},
  \citenamefont {{Babanly}}, \citenamefont {{Mamedov}}, \citenamefont
  {{Abdullayev}}, \citenamefont {{Zverev}}, \citenamefont {{B{\"u}chner}},
  \citenamefont {{Schwier}}, \citenamefont {{Kumar}}, \citenamefont {{Kimura}},
  \citenamefont {{Petaccia}}, \citenamefont {{Di Santo}}, \citenamefont
  {{Vidal}}, \citenamefont {{Schatz}}, \citenamefont {{Ki{\ss}ner}},
  \citenamefont {{Min}}, \citenamefont {{Moser}}, \citenamefont {{Peixoto}},
  \citenamefont {{Reinert}}, \citenamefont {{Ernst}}, \citenamefont
  {{Echenique}}, \citenamefont {{Isaeva}},\ and\ \citenamefont
  {{Chulkov}}}]{otrokov2019}%
  \BibitemOpen
  \bibfield  {author} {\bibinfo {author} {\bibfnamefont {Mikhail~M.}\
  \bibnamefont {{Otrokov}}}, \bibinfo {author} {\bibfnamefont {Ilya~I.}\
  \bibnamefont {{Klimovskikh}}}, \bibinfo {author} {\bibfnamefont {Hendrik}\
  \bibnamefont {{Bentmann}}}, \bibinfo {author} {\bibfnamefont {Alexander}\
  \bibnamefont {{Zeugner}}}, \bibinfo {author} {\bibfnamefont {Ziya~S.}\
  \bibnamefont {{Aliev}}}, \bibinfo {author} {\bibfnamefont {Sebastian}\
  \bibnamefont {{Gass}}}, \bibinfo {author} {\bibfnamefont {Anja U.~B.}\
  \bibnamefont {{Wolter}}}, \bibinfo {author} {\bibfnamefont {Alexand ra~V.}\
  \bibnamefont {{Koroleva}}}, \bibinfo {author} {\bibfnamefont {Dmitry}\
  \bibnamefont {{Estyunin}}}, \bibinfo {author} {\bibfnamefont {Alexander~M.}\
  \bibnamefont {{Shikin}}}, \bibinfo {author} {\bibfnamefont {Mar{\'\i}a}\
  \bibnamefont {{Blanco-Rey}}}, \bibinfo {author} {\bibfnamefont {Martin}\
  \bibnamefont {{Hoffmann}}}, \bibinfo {author} {\bibfnamefont {Alexand
  ra~Yu.}\ \bibnamefont {{Vyazovskaya}}}, \bibinfo {author} {\bibfnamefont
  {Sergey~V.}\ \bibnamefont {{Eremeev}}}, \bibinfo {author} {\bibfnamefont
  {Yury~M.}\ \bibnamefont {{Koroteev}}}, \bibinfo {author} {\bibfnamefont
  {Imamaddin~R.}\ \bibnamefont {{Amiraslanov}}}, \bibinfo {author}
  {\bibfnamefont {Mahammad~B.}\ \bibnamefont {{Babanly}}}, \bibinfo {author}
  {\bibfnamefont {Nazim~T.}\ \bibnamefont {{Mamedov}}}, \bibinfo {author}
  {\bibfnamefont {Nadir~A.}\ \bibnamefont {{Abdullayev}}}, \bibinfo {author}
  {\bibfnamefont {Vladimir~N.}\ \bibnamefont {{Zverev}}}, \bibinfo {author}
  {\bibfnamefont {Bernd}\ \bibnamefont {{B{\"u}chner}}}, \bibinfo {author}
  {\bibfnamefont {Eike~F.}\ \bibnamefont {{Schwier}}}, \bibinfo {author}
  {\bibfnamefont {Shiv}\ \bibnamefont {{Kumar}}}, \bibinfo {author}
  {\bibfnamefont {Akio}\ \bibnamefont {{Kimura}}}, \bibinfo {author}
  {\bibfnamefont {Luca}\ \bibnamefont {{Petaccia}}}, \bibinfo {author}
  {\bibfnamefont {Giovanni}\ \bibnamefont {{Di Santo}}}, \bibinfo {author}
  {\bibfnamefont {Raphael~C.}\ \bibnamefont {{Vidal}}}, \bibinfo {author}
  {\bibfnamefont {Sonja}\ \bibnamefont {{Schatz}}}, \bibinfo {author}
  {\bibfnamefont {Katharina}\ \bibnamefont {{Ki{\ss}ner}}}, \bibinfo {author}
  {\bibfnamefont {Chul-Hee}\ \bibnamefont {{Min}}}, \bibinfo {author}
  {\bibfnamefont {Simon~K.}\ \bibnamefont {{Moser}}}, \bibinfo {author}
  {\bibfnamefont {Thiago R.~F.}\ \bibnamefont {{Peixoto}}}, \bibinfo {author}
  {\bibfnamefont {Friedrich}\ \bibnamefont {{Reinert}}}, \bibinfo {author}
  {\bibfnamefont {Arthur}\ \bibnamefont {{Ernst}}}, \bibinfo {author}
  {\bibfnamefont {Pedro~M.}\ \bibnamefont {{Echenique}}}, \bibinfo {author}
  {\bibfnamefont {Anna}\ \bibnamefont {{Isaeva}}}, \ and\ \bibinfo {author}
  {\bibfnamefont {Evgueni~V.}\ \bibnamefont {{Chulkov}}},\ }\bibfield  {title}
  {\enquote {\bibinfo {title} {Prediction and observation of an
  antiferromagnetic topological insulator},}\ }\href {\doibase
  10.1038/s41586-019-1840-9} {\bibfield  {journal} {\bibinfo  {journal}
  {Nature}\ }\textbf {\bibinfo {volume} {576}},\ \bibinfo {pages} {416--422}
  (\bibinfo {year} {2019})}\BibitemShut {NoStop}%
\bibitem [{\citenamefont {Deng}\ \emph {et~al.}(2020)\citenamefont {Deng},
  \citenamefont {Yu}, \citenamefont {Shi}, \citenamefont {Guo}, \citenamefont
  {Xu}, \citenamefont {Wang}, \citenamefont {Chen},\ and\ \citenamefont
  {Zhang}}]{deng2020}%
  \BibitemOpen
  \bibfield  {author} {\bibinfo {author} {\bibfnamefont {Yujun}\ \bibnamefont
  {Deng}}, \bibinfo {author} {\bibfnamefont {Yijun}\ \bibnamefont {Yu}},
  \bibinfo {author} {\bibfnamefont {Meng~Zhu}\ \bibnamefont {Shi}}, \bibinfo
  {author} {\bibfnamefont {Zhongxun}\ \bibnamefont {Guo}}, \bibinfo {author}
  {\bibfnamefont {Zihan}\ \bibnamefont {Xu}}, \bibinfo {author} {\bibfnamefont
  {Jing}\ \bibnamefont {Wang}}, \bibinfo {author} {\bibfnamefont {Xian~Hui}\
  \bibnamefont {Chen}}, \ and\ \bibinfo {author} {\bibfnamefont {Yuanbo}\
  \bibnamefont {Zhang}},\ }\bibfield  {title} {\enquote {\bibinfo {title}
  {Quantum anomalous hall effect in intrinsic magnetic topological insulator
  mnbi2te4},}\ }\href {\doibase 10.1126/science.aax8156} {\bibfield  {journal}
  {\bibinfo  {journal} {Science}\ }\textbf {\bibinfo {volume} {367}},\ \bibinfo
  {pages} {895--900} (\bibinfo {year} {2020})}\BibitemShut {NoStop}%
\bibitem [{\citenamefont {Liu}\ \emph {et~al.}(2020)\citenamefont {Liu},
  \citenamefont {Wang}, \citenamefont {Li}, \citenamefont {Wu}, \citenamefont
  {Li}, \citenamefont {Li}, \citenamefont {He}, \citenamefont {Xu},
  \citenamefont {Zhang},\ and\ \citenamefont {Wang}}]{liu2020}%
  \BibitemOpen
  \bibfield  {author} {\bibinfo {author} {\bibfnamefont {Chang}\ \bibnamefont
  {Liu}}, \bibinfo {author} {\bibfnamefont {Yongchao}\ \bibnamefont {Wang}},
  \bibinfo {author} {\bibfnamefont {Hao}\ \bibnamefont {Li}}, \bibinfo {author}
  {\bibfnamefont {Yang}\ \bibnamefont {Wu}}, \bibinfo {author} {\bibfnamefont
  {Yaoxin}\ \bibnamefont {Li}}, \bibinfo {author} {\bibfnamefont {Jiaheng}\
  \bibnamefont {Li}}, \bibinfo {author} {\bibfnamefont {Ke}~\bibnamefont {He}},
  \bibinfo {author} {\bibfnamefont {Yong}\ \bibnamefont {Xu}}, \bibinfo
  {author} {\bibfnamefont {Jinsong}\ \bibnamefont {Zhang}}, \ and\ \bibinfo
  {author} {\bibfnamefont {Yayu}\ \bibnamefont {Wang}},\ }\bibfield  {title}
  {\enquote {\bibinfo {title} {Robust axion insulator and chern insulator
  phases in a two-dimensional antiferromagnetic topological insulator},}\
  }\href {\doibase 10.1038/s41563-019-0573-3} {\bibfield  {journal} {\bibinfo
  {journal} {Nature Mat.}\ }\textbf {\bibinfo {volume} {19}},\ \bibinfo {pages}
  {522--527} (\bibinfo {year} {2020})}\BibitemShut {NoStop}%
\bibitem [{\citenamefont {Ge}\ \emph {et~al.}(2020)\citenamefont {Ge},
  \citenamefont {Liu}, \citenamefont {Li}, \citenamefont {Li}, \citenamefont
  {Luo}, \citenamefont {Wu}, \citenamefont {Xu},\ and\ \citenamefont
  {Wang}}]{ge2020}%
  \BibitemOpen
  \bibfield  {author} {\bibinfo {author} {\bibfnamefont {Jun}\ \bibnamefont
  {Ge}}, \bibinfo {author} {\bibfnamefont {Yanzhao}\ \bibnamefont {Liu}},
  \bibinfo {author} {\bibfnamefont {Jiaheng}\ \bibnamefont {Li}}, \bibinfo
  {author} {\bibfnamefont {Hao}\ \bibnamefont {Li}}, \bibinfo {author}
  {\bibfnamefont {Tianchuang}\ \bibnamefont {Luo}}, \bibinfo {author}
  {\bibfnamefont {Yang}\ \bibnamefont {Wu}}, \bibinfo {author} {\bibfnamefont
  {Yong}\ \bibnamefont {Xu}}, \ and\ \bibinfo {author} {\bibfnamefont {Jian}\
  \bibnamefont {Wang}},\ }\bibfield  {title} {\enquote {\bibinfo {title}
  {{High-Chern-number and high-temperature quantum Hall effect without Landau
  levels}},}\ }\href {\doibase 10.1093/nsr/nwaa089} {\bibfield  {journal}
  {\bibinfo  {journal} {National Sci. Rev.}\ }\textbf {\bibinfo {volume} {7}},\
  \bibinfo {pages} {1280--1287} (\bibinfo {year} {2020})}\BibitemShut {NoStop}%
\bibitem [{\citenamefont {Lee}\ \emph {et~al.}(2019)\citenamefont {Lee},
  \citenamefont {Zhu}, \citenamefont {Wang}, \citenamefont {Miao},
  \citenamefont {Pillsbury}, \citenamefont {Yi}, \citenamefont {Kempinger},
  \citenamefont {Hu}, \citenamefont {Heikes}, \citenamefont {Quarterman},
  \citenamefont {Ratcliff}, \citenamefont {Borchers}, \citenamefont {Zhang},
  \citenamefont {Ke}, \citenamefont {Graf}, \citenamefont {Alem}, \citenamefont
  {Chang}, \citenamefont {Samarth},\ and\ \citenamefont {Mao}}]{lee2019}%
  \BibitemOpen
  \bibfield  {author} {\bibinfo {author} {\bibfnamefont {Seng~Huat}\
  \bibnamefont {Lee}}, \bibinfo {author} {\bibfnamefont {Yanglin}\ \bibnamefont
  {Zhu}}, \bibinfo {author} {\bibfnamefont {Yu}~\bibnamefont {Wang}}, \bibinfo
  {author} {\bibfnamefont {Leixin}\ \bibnamefont {Miao}}, \bibinfo {author}
  {\bibfnamefont {Timothy}\ \bibnamefont {Pillsbury}}, \bibinfo {author}
  {\bibfnamefont {Hemian}\ \bibnamefont {Yi}}, \bibinfo {author} {\bibfnamefont
  {Susan}\ \bibnamefont {Kempinger}}, \bibinfo {author} {\bibfnamefont {Jin}\
  \bibnamefont {Hu}}, \bibinfo {author} {\bibfnamefont {Colin~A.}\ \bibnamefont
  {Heikes}}, \bibinfo {author} {\bibfnamefont {P.}~\bibnamefont {Quarterman}},
  \bibinfo {author} {\bibfnamefont {William}\ \bibnamefont {Ratcliff}},
  \bibinfo {author} {\bibfnamefont {Julie~A.}\ \bibnamefont {Borchers}},
  \bibinfo {author} {\bibfnamefont {Heda}\ \bibnamefont {Zhang}}, \bibinfo
  {author} {\bibfnamefont {Xianglin}\ \bibnamefont {Ke}}, \bibinfo {author}
  {\bibfnamefont {David}\ \bibnamefont {Graf}}, \bibinfo {author}
  {\bibfnamefont {Nasim}\ \bibnamefont {Alem}}, \bibinfo {author}
  {\bibfnamefont {Cui-Zu}\ \bibnamefont {Chang}}, \bibinfo {author}
  {\bibfnamefont {Nitin}\ \bibnamefont {Samarth}}, \ and\ \bibinfo {author}
  {\bibfnamefont {Zhiqiang}\ \bibnamefont {Mao}},\ }\bibfield  {title}
  {\enquote {\bibinfo {title} {Spin scattering and noncollinear spin
  structure-induced intrinsic anomalous hall effect in antiferromagnetic
  topological insulator
  $\mathrm{MnB}{\mathrm{i}}_{2}\mathrm{T}{\mathrm{e}}_{4}$},}\ }\href {\doibase
  10.1103/PhysRevResearch.1.012011} {\bibfield  {journal} {\bibinfo  {journal}
  {Phys. Rev. Research}\ }\textbf {\bibinfo {volume} {1}},\ \bibinfo {pages}
  {012011} (\bibinfo {year} {2019})}\BibitemShut {NoStop}%
\bibitem [{\citenamefont {Yan}\ \emph {et~al.}(2019)\citenamefont {Yan},
  \citenamefont {Zhang}, \citenamefont {Heitmann}, \citenamefont {Huang},
  \citenamefont {Chen}, \citenamefont {Cheng}, \citenamefont {Wu},
  \citenamefont {Vaknin}, \citenamefont {Sales},\ and\ \citenamefont
  {McQueeney}}]{yan2019}%
  \BibitemOpen
  \bibfield  {author} {\bibinfo {author} {\bibfnamefont {J.-Q.}\ \bibnamefont
  {Yan}}, \bibinfo {author} {\bibfnamefont {Q.}~\bibnamefont {Zhang}}, \bibinfo
  {author} {\bibfnamefont {T.}~\bibnamefont {Heitmann}}, \bibinfo {author}
  {\bibfnamefont {Z.}~\bibnamefont {Huang}}, \bibinfo {author} {\bibfnamefont
  {K.~Y.}\ \bibnamefont {Chen}}, \bibinfo {author} {\bibfnamefont {J.-G.}\
  \bibnamefont {Cheng}}, \bibinfo {author} {\bibfnamefont {W.}~\bibnamefont
  {Wu}}, \bibinfo {author} {\bibfnamefont {D.}~\bibnamefont {Vaknin}}, \bibinfo
  {author} {\bibfnamefont {B.~C.}\ \bibnamefont {Sales}}, \ and\ \bibinfo
  {author} {\bibfnamefont {R.~J.}\ \bibnamefont {McQueeney}},\ }\bibfield
  {title} {\enquote {\bibinfo {title} {Crystal growth and magnetic structure of
  ${\mathrm{mnbi}}_{2}{\mathrm{te}}_{4}$},}\ }\href {\doibase
  10.1103/PhysRevMaterials.3.064202} {\bibfield  {journal} {\bibinfo  {journal}
  {Phys. Rev. Materials}\ }\textbf {\bibinfo {volume} {3}},\ \bibinfo {pages}
  {064202} (\bibinfo {year} {2019})}\BibitemShut {NoStop}%
\bibitem [{\citenamefont {Hao}\ \emph {et~al.}(2019)\citenamefont {Hao},
  \citenamefont {Liu}, \citenamefont {Feng}, \citenamefont {Ma}, \citenamefont
  {Schwier}, \citenamefont {Arita}, \citenamefont {Kumar}, \citenamefont {Hu},
  \citenamefont {Lu}, \citenamefont {Zeng}, \citenamefont {Wang}, \citenamefont
  {Hao}, \citenamefont {Sun}, \citenamefont {Zhang}, \citenamefont {Mei},
  \citenamefont {Ni}, \citenamefont {Wu}, \citenamefont {Shimada},
  \citenamefont {Chen}, \citenamefont {Liu},\ and\ \citenamefont
  {Liu}}]{hao2019}%
  \BibitemOpen
  \bibfield  {author} {\bibinfo {author} {\bibfnamefont {Yu-Jie}\ \bibnamefont
  {Hao}}, \bibinfo {author} {\bibfnamefont {Pengfei}\ \bibnamefont {Liu}},
  \bibinfo {author} {\bibfnamefont {Yue}\ \bibnamefont {Feng}}, \bibinfo
  {author} {\bibfnamefont {Xiao-Ming}\ \bibnamefont {Ma}}, \bibinfo {author}
  {\bibfnamefont {Eike~F.}\ \bibnamefont {Schwier}}, \bibinfo {author}
  {\bibfnamefont {Masashi}\ \bibnamefont {Arita}}, \bibinfo {author}
  {\bibfnamefont {Shiv}\ \bibnamefont {Kumar}}, \bibinfo {author}
  {\bibfnamefont {Chaowei}\ \bibnamefont {Hu}}, \bibinfo {author}
  {\bibfnamefont {Rui'e}\ \bibnamefont {Lu}}, \bibinfo {author} {\bibfnamefont
  {Meng}\ \bibnamefont {Zeng}}, \bibinfo {author} {\bibfnamefont {Yuan}\
  \bibnamefont {Wang}}, \bibinfo {author} {\bibfnamefont {Zhanyang}\
  \bibnamefont {Hao}}, \bibinfo {author} {\bibfnamefont {Hong-Yi}\ \bibnamefont
  {Sun}}, \bibinfo {author} {\bibfnamefont {Ke}~\bibnamefont {Zhang}}, \bibinfo
  {author} {\bibfnamefont {Jiawei}\ \bibnamefont {Mei}}, \bibinfo {author}
  {\bibfnamefont {Ni}~\bibnamefont {Ni}}, \bibinfo {author} {\bibfnamefont
  {Liusuo}\ \bibnamefont {Wu}}, \bibinfo {author} {\bibfnamefont {Kenya}\
  \bibnamefont {Shimada}}, \bibinfo {author} {\bibfnamefont {Chaoyu}\
  \bibnamefont {Chen}}, \bibinfo {author} {\bibfnamefont {Qihang}\ \bibnamefont
  {Liu}}, \ and\ \bibinfo {author} {\bibfnamefont {Chang}\ \bibnamefont
  {Liu}},\ }\bibfield  {title} {\enquote {\bibinfo {title} {Gapless surface
  dirac cone in antiferromagnetic topological insulator
  ${\mathrm{mnbi}}_{2}{\mathrm{te}}_{4}$},}\ }\href {\doibase
  10.1103/PhysRevX.9.041038} {\bibfield  {journal} {\bibinfo  {journal} {Phys.
  Rev. X}\ }\textbf {\bibinfo {volume} {9}},\ \bibinfo {pages} {041038}
  (\bibinfo {year} {2019})}\BibitemShut {NoStop}%
\bibitem [{\citenamefont {Li}\ \emph {et~al.}(2019{\natexlab{b}})\citenamefont
  {Li}, \citenamefont {Gao}, \citenamefont {Duan}, \citenamefont {Xu},
  \citenamefont {Zhu}, \citenamefont {Tian}, \citenamefont {Gao}, \citenamefont
  {Fan}, \citenamefont {Rao}, \citenamefont {Huang}, \citenamefont {Li},
  \citenamefont {Yan}, \citenamefont {Liu}, \citenamefont {Liu}, \citenamefont
  {Huang}, \citenamefont {Li}, \citenamefont {Liu}, \citenamefont {Zhang},
  \citenamefont {Zhang}, \citenamefont {Kondo}, \citenamefont {Shin},
  \citenamefont {Lei}, \citenamefont {Shi}, \citenamefont {Zhang},
  \citenamefont {Weng}, \citenamefont {Qian},\ and\ \citenamefont
  {Ding}}]{lih2019}%
  \BibitemOpen
  \bibfield  {author} {\bibinfo {author} {\bibfnamefont {Hang}\ \bibnamefont
  {Li}}, \bibinfo {author} {\bibfnamefont {Shun-Ye}\ \bibnamefont {Gao}},
  \bibinfo {author} {\bibfnamefont {Shao-Feng}\ \bibnamefont {Duan}}, \bibinfo
  {author} {\bibfnamefont {Yuan-Feng}\ \bibnamefont {Xu}}, \bibinfo {author}
  {\bibfnamefont {Ke-Jia}\ \bibnamefont {Zhu}}, \bibinfo {author}
  {\bibfnamefont {Shang-Jie}\ \bibnamefont {Tian}}, \bibinfo {author}
  {\bibfnamefont {Jia-Cheng}\ \bibnamefont {Gao}}, \bibinfo {author}
  {\bibfnamefont {Wen-Hui}\ \bibnamefont {Fan}}, \bibinfo {author}
  {\bibfnamefont {Zhi-Cheng}\ \bibnamefont {Rao}}, \bibinfo {author}
  {\bibfnamefont {Jie-Rui}\ \bibnamefont {Huang}}, \bibinfo {author}
  {\bibfnamefont {Jia-Jun}\ \bibnamefont {Li}}, \bibinfo {author}
  {\bibfnamefont {Da-Yu}\ \bibnamefont {Yan}}, \bibinfo {author} {\bibfnamefont
  {Zheng-Tai}\ \bibnamefont {Liu}}, \bibinfo {author} {\bibfnamefont
  {Wan-Ling}\ \bibnamefont {Liu}}, \bibinfo {author} {\bibfnamefont {Yao-Bo}\
  \bibnamefont {Huang}}, \bibinfo {author} {\bibfnamefont {Yu-Liang}\
  \bibnamefont {Li}}, \bibinfo {author} {\bibfnamefont {Yi}~\bibnamefont
  {Liu}}, \bibinfo {author} {\bibfnamefont {Guo-Bin}\ \bibnamefont {Zhang}},
  \bibinfo {author} {\bibfnamefont {Peng}\ \bibnamefont {Zhang}}, \bibinfo
  {author} {\bibfnamefont {Takeshi}\ \bibnamefont {Kondo}}, \bibinfo {author}
  {\bibfnamefont {Shik}\ \bibnamefont {Shin}}, \bibinfo {author} {\bibfnamefont
  {He-Chang}\ \bibnamefont {Lei}}, \bibinfo {author} {\bibfnamefont {You-Guo}\
  \bibnamefont {Shi}}, \bibinfo {author} {\bibfnamefont {Wen-Tao}\ \bibnamefont
  {Zhang}}, \bibinfo {author} {\bibfnamefont {Hong-Ming}\ \bibnamefont {Weng}},
  \bibinfo {author} {\bibfnamefont {Tian}\ \bibnamefont {Qian}}, \ and\
  \bibinfo {author} {\bibfnamefont {Hong}\ \bibnamefont {Ding}},\ }\bibfield
  {title} {\enquote {\bibinfo {title} {Dirac surface states in intrinsic
  magnetic topological insulators ${\mathrm{eusn}}_{2}{\mathrm{as}}_{2}$ and
  ${\mathrm{mnbi}}_{2n}{\mathrm{te}}_{3n+1}$},}\ }\href {\doibase
  10.1103/PhysRevX.9.041039} {\bibfield  {journal} {\bibinfo  {journal} {Phys.
  Rev. X}\ }\textbf {\bibinfo {volume} {9}},\ \bibinfo {pages} {041039}
  (\bibinfo {year} {2019}{\natexlab{b}})}\BibitemShut {NoStop}%
\bibitem [{\citenamefont {Chen}\ \emph {et~al.}(2019)\citenamefont {Chen},
  \citenamefont {Xu}, \citenamefont {Li}, \citenamefont {Li}, \citenamefont
  {Wang}, \citenamefont {Zhang}, \citenamefont {Li}, \citenamefont {Wu},
  \citenamefont {Liang}, \citenamefont {Chen}, \citenamefont {Jung},
  \citenamefont {Cacho}, \citenamefont {Mao}, \citenamefont {Liu},
  \citenamefont {Wang}, \citenamefont {Guo}, \citenamefont {Xu}, \citenamefont
  {Liu}, \citenamefont {Yang},\ and\ \citenamefont {Chen}}]{chen2019}%
  \BibitemOpen
  \bibfield  {author} {\bibinfo {author} {\bibfnamefont {Y.~J.}\ \bibnamefont
  {Chen}}, \bibinfo {author} {\bibfnamefont {L.~X.}\ \bibnamefont {Xu}},
  \bibinfo {author} {\bibfnamefont {J.~H.}\ \bibnamefont {Li}}, \bibinfo
  {author} {\bibfnamefont {Y.~W.}\ \bibnamefont {Li}}, \bibinfo {author}
  {\bibfnamefont {H.~Y.}\ \bibnamefont {Wang}}, \bibinfo {author}
  {\bibfnamefont {C.~F.}\ \bibnamefont {Zhang}}, \bibinfo {author}
  {\bibfnamefont {H.}~\bibnamefont {Li}}, \bibinfo {author} {\bibfnamefont
  {Y.}~\bibnamefont {Wu}}, \bibinfo {author} {\bibfnamefont {A.~J.}\
  \bibnamefont {Liang}}, \bibinfo {author} {\bibfnamefont {C.}~\bibnamefont
  {Chen}}, \bibinfo {author} {\bibfnamefont {S.~W.}\ \bibnamefont {Jung}},
  \bibinfo {author} {\bibfnamefont {C.}~\bibnamefont {Cacho}}, \bibinfo
  {author} {\bibfnamefont {Y.~H.}\ \bibnamefont {Mao}}, \bibinfo {author}
  {\bibfnamefont {S.}~\bibnamefont {Liu}}, \bibinfo {author} {\bibfnamefont
  {M.~X.}\ \bibnamefont {Wang}}, \bibinfo {author} {\bibfnamefont {Y.~F.}\
  \bibnamefont {Guo}}, \bibinfo {author} {\bibfnamefont {Y.}~\bibnamefont
  {Xu}}, \bibinfo {author} {\bibfnamefont {Z.~K.}\ \bibnamefont {Liu}},
  \bibinfo {author} {\bibfnamefont {L.~X.}\ \bibnamefont {Yang}}, \ and\
  \bibinfo {author} {\bibfnamefont {Y.~L.}\ \bibnamefont {Chen}},\ }\bibfield
  {title} {\enquote {\bibinfo {title} {Topological electronic structure and its
  temperature evolution in antiferromagnetic topological insulator
  ${\mathrm{mnbi}}_{2}{\mathrm{te}}_{4}$},}\ }\href {\doibase
  10.1103/PhysRevX.9.041040} {\bibfield  {journal} {\bibinfo  {journal} {Phys.
  Rev. X}\ }\textbf {\bibinfo {volume} {9}},\ \bibinfo {pages} {041040}
  (\bibinfo {year} {2019})}\BibitemShut {NoStop}%
\bibitem [{\citenamefont {Zhang}\ \emph {et~al.}(2020)\citenamefont {Zhang},
  \citenamefont {Wang}, \citenamefont {Shi}, \citenamefont {Zhu}, \citenamefont
  {Zhang},\ and\ \citenamefont {Wang}}]{zhang2020}%
  \BibitemOpen
  \bibfield  {author} {\bibinfo {author} {\bibfnamefont {Jinlong}\ \bibnamefont
  {Zhang}}, \bibinfo {author} {\bibfnamefont {Dinghui}\ \bibnamefont {Wang}},
  \bibinfo {author} {\bibfnamefont {Minji}\ \bibnamefont {Shi}}, \bibinfo
  {author} {\bibfnamefont {Tongshuai}\ \bibnamefont {Zhu}}, \bibinfo {author}
  {\bibfnamefont {Haijun}\ \bibnamefont {Zhang}}, \ and\ \bibinfo {author}
  {\bibfnamefont {Jing}\ \bibnamefont {Wang}},\ }\bibfield  {title} {\enquote
  {\bibinfo {title} {Large dynamicak axion field in topological
  antiferromagnetic insulator mn2bi2te5},}\ }\href {\doibase
  10.1088/0256-307X/37/7/077304} {\bibfield  {journal} {\bibinfo  {journal}
  {Chin. Phys. Lett.}\ }\textbf {\bibinfo {volume} {37}},\ \bibinfo {eid}
  {077304} (\bibinfo {year} {2020})}\BibitemShut {NoStop}%
\bibitem [{lv2()}]{lv2020}%
  \BibitemOpen
  \href@noop {} {}\bibinfo {note} {Y. Lv $\emph{et al}$., to be published
  (2020).}\BibitemShut {Stop}%
\bibitem [{\citenamefont {Hou}\ and\ \citenamefont {Wu}(2019)}]{hou2019}%
  \BibitemOpen
  \bibfield  {author} {\bibinfo {author} {\bibfnamefont {Yusheng}\ \bibnamefont
  {Hou}}\ and\ \bibinfo {author} {\bibfnamefont {Ruqian}\ \bibnamefont {Wu}},\
  }\bibfield  {title} {\enquote {\bibinfo {title} {Axion insulator state in a
  ferromagnet/topological insulator/antiferromagnet heterostructure},}\ }\href
  {\doibase 10.1021/acs.nanolett.9b00047} {\bibfield  {journal} {\bibinfo
  {journal} {Nano Lett.}\ }\textbf {\bibinfo {volume} {19}},\ \bibinfo {pages}
  {2472--2477} (\bibinfo {year} {2019})}\BibitemShut {NoStop}%
\bibitem [{sup()}]{supplementary}%
  \BibitemOpen
  \href@noop {} {}\bibinfo {note} {See Supplemental Material at [url], for
  technical details on first-principles calculations and effective models with
  fitting parameters, which includes
  Refs.~\cite{Blochl1994,Kresse1996,Perdew1996,grimme2010,dudarev1998,mostofi2008,Wu2017}.}\BibitemShut
  {Stop}%
\bibitem [{\citenamefont {Zhang}\ \emph {et~al.}(2009)\citenamefont {Zhang},
  \citenamefont {Liu}, \citenamefont {Qi}, \citenamefont {Dai}, \citenamefont
  {Fang},\ and\ \citenamefont {Zhang}}]{zhang2009}%
  \BibitemOpen
  \bibfield  {author} {\bibinfo {author} {\bibfnamefont {Haijun}\ \bibnamefont
  {Zhang}}, \bibinfo {author} {\bibfnamefont {Chao-Xing}\ \bibnamefont {Liu}},
  \bibinfo {author} {\bibfnamefont {Xiao-Liang}\ \bibnamefont {Qi}}, \bibinfo
  {author} {\bibfnamefont {Xi}~\bibnamefont {Dai}}, \bibinfo {author}
  {\bibfnamefont {Zhong}\ \bibnamefont {Fang}}, \ and\ \bibinfo {author}
  {\bibfnamefont {Shou-Cheng}\ \bibnamefont {Zhang}},\ }\bibfield  {title}
  {\enquote {\bibinfo {title} {{Topological insulators in
  {$\mathrm{Bi_2Se_3}$}, {$\mathrm{Bi_2Te_3}$} and {$\mathrm{Sb_2Te_3}$} with a
  single Dirac cone on the surface}},}\ }\href@noop {} {\bibfield  {journal}
  {\bibinfo  {journal} {Nature Phys.}\ }\textbf {\bibinfo {volume} {5}},\
  \bibinfo {pages} {438} (\bibinfo {year} {2009})}\BibitemShut {NoStop}%
\bibitem [{\citenamefont {Burkov}\ and\ \citenamefont
  {Balents}(2011)}]{burkov2011}%
  \BibitemOpen
  \bibfield  {author} {\bibinfo {author} {\bibfnamefont {A.~A.}\ \bibnamefont
  {Burkov}}\ and\ \bibinfo {author} {\bibfnamefont {Leon}\ \bibnamefont
  {Balents}},\ }\bibfield  {title} {\enquote {\bibinfo {title} {Weyl semimetal
  in a topological insulator multilayer},}\ }\href {\doibase
  10.1103/PhysRevLett.107.127205} {\bibfield  {journal} {\bibinfo  {journal}
  {Phys. Rev. Lett.}\ }\textbf {\bibinfo {volume} {107}},\ \bibinfo {pages}
  {127205} (\bibinfo {year} {2011})}\BibitemShut {NoStop}%
\bibitem [{\citenamefont {Liu}\ \emph {et~al.}(2010)\citenamefont {Liu},
  \citenamefont {Zhang}, \citenamefont {Yan}, \citenamefont {Qi}, \citenamefont
  {Frauenheim}, \citenamefont {Dai}, \citenamefont {Fang},\ and\ \citenamefont
  {Zhang}}]{liu2010a}%
  \BibitemOpen
  \bibfield  {author} {\bibinfo {author} {\bibfnamefont {Chao-Xing}\
  \bibnamefont {Liu}}, \bibinfo {author} {\bibfnamefont {Hai-Jun}\ \bibnamefont
  {Zhang}}, \bibinfo {author} {\bibfnamefont {Binghai}\ \bibnamefont {Yan}},
  \bibinfo {author} {\bibfnamefont {Xiao-Liang}\ \bibnamefont {Qi}}, \bibinfo
  {author} {\bibfnamefont {Thomas}\ \bibnamefont {Frauenheim}}, \bibinfo
  {author} {\bibfnamefont {Xi}~\bibnamefont {Dai}}, \bibinfo {author}
  {\bibfnamefont {Zhong}\ \bibnamefont {Fang}}, \ and\ \bibinfo {author}
  {\bibfnamefont {Shou-Cheng}\ \bibnamefont {Zhang}},\ }\bibfield  {title}
  {\enquote {\bibinfo {title} {Oscillatory crossover from two-dimensional to
  three-dimensional topological insulators},}\ }\href {\doibase
  10.1103/PhysRevB.81.041307} {\bibfield  {journal} {\bibinfo  {journal} {Phys.
  Rev. B}\ }\textbf {\bibinfo {volume} {81}},\ \bibinfo {pages} {041307}
  (\bibinfo {year} {2010})}\BibitemShut {NoStop}%
\bibitem [{\citenamefont {Liu}\ and\ \citenamefont {Wang}(2020)}]{liuzc2020}%
  \BibitemOpen
  \bibfield  {author} {\bibinfo {author} {\bibfnamefont {Zhaochen}\
  \bibnamefont {Liu}}\ and\ \bibinfo {author} {\bibfnamefont {Jing}\
  \bibnamefont {Wang}},\ }\bibfield  {title} {\enquote {\bibinfo {title}
  {Anisotropic topological magnetoelectric effect in axion insulators},}\
  }\href {\doibase 10.1103/PhysRevB.101.205130} {\bibfield  {journal} {\bibinfo
   {journal} {Phys. Rev. B}\ }\textbf {\bibinfo {volume} {101}},\ \bibinfo
  {pages} {205130} (\bibinfo {year} {2020})}\BibitemShut {NoStop}%
\bibitem [{\citenamefont {Lian}\ \emph {et~al.}(2020)\citenamefont {Lian},
  \citenamefont {Liu}, \citenamefont {Zhang},\ and\ \citenamefont
  {Wang}}]{lian2020}%
  \BibitemOpen
  \bibfield  {author} {\bibinfo {author} {\bibfnamefont {Biao}\ \bibnamefont
  {Lian}}, \bibinfo {author} {\bibfnamefont {Zhaochen}\ \bibnamefont {Liu}},
  \bibinfo {author} {\bibfnamefont {Yuanbo}\ \bibnamefont {Zhang}}, \ and\
  \bibinfo {author} {\bibfnamefont {Jing}\ \bibnamefont {Wang}},\ }\bibfield
  {title} {\enquote {\bibinfo {title} {Flat chern band from twisted bilayer
  ${\mathrm{mnbi}}_{2}{\mathrm{te}}_{4}$},}\ }\href {\doibase
  10.1103/PhysRevLett.124.126402} {\bibfield  {journal} {\bibinfo  {journal}
  {Phys. Rev. Lett.}\ }\textbf {\bibinfo {volume} {124}},\ \bibinfo {pages}
  {126402} (\bibinfo {year} {2020})}\BibitemShut {NoStop}%
\bibitem [{\citenamefont {Bl\"ochl}(1994)}]{Blochl1994}%
  \BibitemOpen
  \bibfield  {author} {\bibinfo {author} {\bibfnamefont {P.~E.}\ \bibnamefont
  {Bl\"ochl}},\ }\bibfield  {title} {\enquote {\bibinfo {title} {Projector
  augmented-wave method},}\ }\href {\doibase 10.1103/PhysRevB.50.17953}
  {\bibfield  {journal} {\bibinfo  {journal} {Phys. Rev. B}\ }\textbf {\bibinfo
  {volume} {50}},\ \bibinfo {pages} {17953--17979} (\bibinfo {year}
  {1994})}\BibitemShut {NoStop}%
\bibitem [{\citenamefont {Kresse}\ and\ \citenamefont
  {Furthm\"uller}(1996)}]{Kresse1996}%
  \BibitemOpen
  \bibfield  {author} {\bibinfo {author} {\bibfnamefont {G.}~\bibnamefont
  {Kresse}}\ and\ \bibinfo {author} {\bibfnamefont {J.}~\bibnamefont
  {Furthm\"uller}},\ }\bibfield  {title} {\enquote {\bibinfo {title} {Efficient
  iterative schemes for ab initio total-energy calculations using a plane-wave
  basis set},}\ }\href {\doibase 10.1103/PhysRevB.54.11169} {\bibfield
  {journal} {\bibinfo  {journal} {Phys. Rev. B}\ }\textbf {\bibinfo {volume}
  {54}},\ \bibinfo {pages} {11169--11186} (\bibinfo {year} {1996})}\BibitemShut
  {NoStop}%
\bibitem [{\citenamefont {Perdew}\ \emph {et~al.}(1996)\citenamefont {Perdew},
  \citenamefont {Burke},\ and\ \citenamefont {Ernzerhof}}]{Perdew1996}%
  \BibitemOpen
  \bibfield  {author} {\bibinfo {author} {\bibfnamefont {John~P.}\ \bibnamefont
  {Perdew}}, \bibinfo {author} {\bibfnamefont {Kieron}\ \bibnamefont {Burke}},
  \ and\ \bibinfo {author} {\bibfnamefont {Matthias}\ \bibnamefont
  {Ernzerhof}},\ }\bibfield  {title} {\enquote {\bibinfo {title} {Generalized
  gradient approximation made simple},}\ }\href {\doibase
  10.1103/PhysRevLett.77.3865} {\bibfield  {journal} {\bibinfo  {journal}
  {Phys. Rev. Lett.}\ }\textbf {\bibinfo {volume} {77}},\ \bibinfo {pages}
  {3865--3868} (\bibinfo {year} {1996})}\BibitemShut {NoStop}%
\bibitem [{\citenamefont {Grimme}\ \emph {et~al.}(2010)\citenamefont {Grimme},
  \citenamefont {Antony}, \citenamefont {Ehrlich},\ and\ \citenamefont
  {Krieg}}]{grimme2010}%
  \BibitemOpen
  \bibfield  {author} {\bibinfo {author} {\bibfnamefont {Stefan}\ \bibnamefont
  {Grimme}}, \bibinfo {author} {\bibfnamefont {Jens}\ \bibnamefont {Antony}},
  \bibinfo {author} {\bibfnamefont {Stephan}\ \bibnamefont {Ehrlich}}, \ and\
  \bibinfo {author} {\bibfnamefont {Helge}\ \bibnamefont {Krieg}},\ }\bibfield
  {title} {\enquote {\bibinfo {title} {A consistent and accurate ab initio
  parametrization of density functional dispersion correction (dft-d) for the
  94 elements h-pu},}\ }\href@noop {} {\bibfield  {journal} {\bibinfo
  {journal} {J. Chem. Phys.}\ }\textbf {\bibinfo {volume} {132}},\ \bibinfo
  {pages} {154104} (\bibinfo {year} {2010})}\BibitemShut {NoStop}%
\bibitem [{\citenamefont {Dudarev}\ \emph {et~al.}(1998)\citenamefont
  {Dudarev}, \citenamefont {Botton}, \citenamefont {Savrasov}, \citenamefont
  {Humphreys},\ and\ \citenamefont {Sutton}}]{dudarev1998}%
  \BibitemOpen
  \bibfield  {author} {\bibinfo {author} {\bibfnamefont {S.~L.}\ \bibnamefont
  {Dudarev}}, \bibinfo {author} {\bibfnamefont {G.~A.}\ \bibnamefont {Botton}},
  \bibinfo {author} {\bibfnamefont {S.~Y.}\ \bibnamefont {Savrasov}}, \bibinfo
  {author} {\bibfnamefont {C.~J.}\ \bibnamefont {Humphreys}}, \ and\ \bibinfo
  {author} {\bibfnamefont {A.~P.}\ \bibnamefont {Sutton}},\ }\bibfield  {title}
  {\enquote {\bibinfo {title} {Electron-energy-loss spectra and the structural
  stability of nickel oxide: An lsda+u study},}\ }\href {\doibase
  10.1103/PhysRevB.57.1505} {\bibfield  {journal} {\bibinfo  {journal} {Phys.
  Rev. B}\ }\textbf {\bibinfo {volume} {57}},\ \bibinfo {pages} {1505--1509}
  (\bibinfo {year} {1998})}\BibitemShut {NoStop}%
\bibitem [{\citenamefont {Mostofi}\ \emph {et~al.}(2008)\citenamefont
  {Mostofi}, \citenamefont {Yates}, \citenamefont {Lee}, \citenamefont {Souza},
  \citenamefont {Vanderbilt},\ and\ \citenamefont {Marzari}}]{mostofi2008}%
  \BibitemOpen
  \bibfield  {author} {\bibinfo {author} {\bibfnamefont {Arash~A}\ \bibnamefont
  {Mostofi}}, \bibinfo {author} {\bibfnamefont {Jonathan~R}\ \bibnamefont
  {Yates}}, \bibinfo {author} {\bibfnamefont {Young-Su}\ \bibnamefont {Lee}},
  \bibinfo {author} {\bibfnamefont {Ivo}\ \bibnamefont {Souza}}, \bibinfo
  {author} {\bibfnamefont {David}\ \bibnamefont {Vanderbilt}}, \ and\ \bibinfo
  {author} {\bibfnamefont {Nicola}\ \bibnamefont {Marzari}},\ }\bibfield
  {title} {\enquote {\bibinfo {title} {wannier90: A tool for obtaining
  maximally-localised wannier functions},}\ }\href@noop {} {\bibfield
  {journal} {\bibinfo  {journal} {Comput. Phys. Commun}\ }\textbf {\bibinfo
  {volume} {178}},\ \bibinfo {pages} {685--699} (\bibinfo {year}
  {2008})}\BibitemShut {NoStop}%
\bibitem [{\citenamefont {Wu}\ \emph {et~al.}(2018)\citenamefont {Wu},
  \citenamefont {Zhang}, \citenamefont {Song}, \citenamefont {Troyer},\ and\
  \citenamefont {Soluyanov}}]{Wu2017}%
  \BibitemOpen
  \bibfield  {author} {\bibinfo {author} {\bibfnamefont {QuanSheng}\
  \bibnamefont {Wu}}, \bibinfo {author} {\bibfnamefont {ShengNan}\ \bibnamefont
  {Zhang}}, \bibinfo {author} {\bibfnamefont {Hai-Feng}\ \bibnamefont {Song}},
  \bibinfo {author} {\bibfnamefont {Matthias}\ \bibnamefont {Troyer}}, \ and\
  \bibinfo {author} {\bibfnamefont {Alexey~A.}\ \bibnamefont {Soluyanov}},\
  }\bibfield  {title} {\enquote {\bibinfo {title} {Wanniertools : An
  open-source software package for novel topological materials},}\ }\href
  {\doibase https://doi.org/10.1016/j.cpc.2017.09.033} {\bibfield  {journal}
  {\bibinfo  {journal} {Comput. Phys. Commun}\ }\textbf {\bibinfo {volume}
  {224}},\ \bibinfo {pages} {405 -- 416} (\bibinfo {year} {2018})}\BibitemShut
  {NoStop}%
\end{thebibliography}
\end{document}